\documentclass[12pt]{article}

\usepackage{graphics,epsfig}
\usepackage{amsbsy}
\usepackage{amsfonts}
\usepackage{amsmath}
\usepackage{graphicx}

\def\IR{\mathbb{R}}
\def\IC{\mathbb{C}}
\def\IZ{\mathbb{Z}}
\def\IP{\mathbb{P}}

\def\id{\protect{{1 \kern-.28em {\rm l}}}}

\def\k{\kappa}

\def\p{{\partial}}
\def\nn{\nonumber}

\def \bs {\bigskip}

\def\dalemb#1#2{{\vbox{\hrule height .#2pt
        \hbox{\vrule width.#2pt height#1pt \kern#1pt
                \vrule width.#2pt}
        \hrule height.#2pt}}}

\def\half{{\textstyle{1\over2}}}
\let\a=\alpha \let\b=\beta \let\g=\gamma \let\d=\delta \let\e=\epsilon
\let\z=\zeta  \let\th=\theta  \let\k=\kappa
\let\l=\lambda \let\m=\mu  \let\x=\xi \let\p=\pi 
\let\s=\sigma \let\t=\tau   \let\c=\chi 
 \let\vep=\varepsilon
\let\w=\omega      \let\G=\Gamma \let\D=\Delta \let\Th=\Theta \let\L=\Lambda
 \let\P=\Pi \let\S=\Sigma  
\let\C=\Chi \let\W=\Omega
\let\la=\label \let\ci=\cite 
  
\def\nn{\nonumber} \def\bd{\begin{document}} \def\ed{\end{document}}
\def\ds{\documentstyle} \let\fr=\frac \let\bl=\bigl \let\br=\bigr
\let\Br=\Bigr \let\Bl=\Bigl
\let\bm=\bibitem
\let\na=\nabla
\def\tU{{\widetilde U}}
\let\pa=\partial \let\ov=\overline
\def\ie{{\it i.e.\ }}
\newcommand{\be}{\begin{equation}}
\newcommand{\ee}{\end{equation}}
\def\ba{\begin{array}}
\def\ea{\end{array}}
\def\ft#1#2{{\textstyle{{\scriptstyle #1}\over {\scriptstyle #2}}}}
\def\fft#1#2{{#1 \over #2}}
\def\F#1#2{{ F_{#1}^{(#2)} }}
\def\cF#1#2{{ {\cal F}_{#1}^{(#2)} }}

\def\={\, =\, }
\def\+{\, +\, }
\def\-{\, -\, }

\def\R{{\bf R}}
\def\sst#1{{\scriptscriptstyle #1}}
\def\oneone{\rlap 1\mkern4mu{\rm l}}
\def\e7{E_{7(+7)}}
\def\td{\tilde}
\def\wtd{\widetilde}
\def\im{{\rm i}}
\newcommand{\ho}[1]{$\, ^{#1}$}
\newcommand{\hoch}[1]{$\, ^{#1}$}
\newcommand{\bea}{\begin{eqnarray}}
\newcommand{\eea}{\end{eqnarray}}
\newcommand{\ra}{\rightarrow}
\newcommand{\lra}{\longrightarrow}
\newcommand{\Lra}{\Leftrightarrow}
\newcommand{\ap}{\alpha^\prime}
\newcommand{\bp}{\tilde \beta^\prime}
\newcommand{\cB}{{\cal B}}
\newcommand{\cO}{{\cal O}}
\newcommand{\vecx}{\vec{x}}
\newcommand{\vecy}{\vec{y}}
\newcommand{\vecp}{\vec{p}}
\newcommand{\vecq}{\vec{q}}
\newcommand{\tr}{{\rm tr} }
\newcommand{\Tr}{{\rm Tr} }

\newcommand{\cL}{{\cal L}}
\newcommand{\cA}{{\cal A}}
\newcommand{\cD}{{\cal D}}
\def\sst#1{{\scriptscriptstyle #1}}

\def\ve{\varepsilon}
\def\vf{\varphi}
\def\F{\Phi}
\def\wg{\wedge}

\def \foot {\footnote}
\def \bi{\bibitem}

\def \tr {{\rm tr}}
\def \ha {{1 \over 2}}
\def \td {\tilde}
\def \ci{\cite}
\def \N {{\mathcal N}}
\def \ww {\Omega}
\def \const {{\rm const}}
\def \ss {\sum_{i=1}^3 }
\def \t {\tau}
\def\S{{\mathcal S} }
\def \nn {\nu}
\def \XX {{\rm X}}

\def \lra {\leftrightarrow}
\def \vom {{\bar \omega}}
\def \E {{\mathcal  E}} \def \J {{\mathcal  J}}
\def \YY {{\rm Y}}

\def \d {\del}
\def \rJ {{J}}
\def \sms {sigma models\ }
\def \sm {sigma model\ }
\def \L {\Lambda}
\def \gl {\ell}
\def \tr {{\rm tr\ }}
\def\z{\zeta}
\def\zi{\zeta_1}
\def\zii{\zeta_2}
\def\K{\mbox{K}}
\def\eE{\mbox{E}}   \def \vt {\vartheta}
\def \vr {\varrho}
\def \wup {w}

\def\dg{\dagger}
\def\a{\alpha}
\def\b{\beta}
\def\e{\varepsilon}
\def\p{\phi}
\def\ap{\alpha^\prime}
\def\I{{\cal I}}

\def\R{{\bf R}}
\def\Z{{\bf Z}}
\def\C{{\bf C}}
\def\P{{\bf P}}
\def\xb{{\bar X}}
\def\Tr{{\rm  Tr}}
\def\tr{{\rm  tr}}

\def \del{\partial}
\def \a {\alpha}
\def \aa {{\a'}}
\def\g{\gamma}
\def\s{\sigma}
\def\z{\zeta}
\def\zi{\zeta_1}
\def\zii{\zeta_2}
\def\ov{\over}

\def\I{{\cal I}}
\def\J{{\mathcal J}}
\def \ok {{1\ov \k}}
\def\LL{{\mathcal L }}
\def \jL {{J}}
\def \om {\omega}
\def \cL {{\mathcal L}} \def \cH {{\mathcal H}}
\def\E{{\mathcal E}}
\def\w{\omega}
\def\b{\beta}
\def\l{\lambda}
\def\eps{\epsilon}
\def\vep{\varepsilon}
\def \De {{\mathcal D}}

 \def \cV {{\cal V}}
\def  \Jt {  {J}_{\rm tot}    }

\def \k {\kappa}
\def\foot{\footnote}
\def \four{{\textstyle {1\ov 4}}}
 \def \third { \textstyle {1\ov 3
}}
\def\det{\hbox{det}}
\def \ci {\cite}

\def \foot {\footnote}
\def \bi{\bibitem}

\def \tr {{\rm tr}}
\def \ha {{1 \over 2}}
\def \tid {\tilde}
\def \vv {{\rm v}}
\def \tl {{\tilde \l}}
\def \XX {{\rm X}}
\def \ta {{\tilde \a}}
\def \fo { {1\ov 4}}
\def \ep {\epsilon}
\def \inti {{\int^{2\pi}_0 {d \sigma \ov 2 \pi}}}

\def \d {\partial}
\def \K {{\rm S}}
\def \el {\ell}
\def \Tr {{\rm Tr}}
\def \P {\Phi}
\def \l  {\lambda}
\def \tl {{\tilde \l}}
\def \bl {{\tilde \l}}
\def \const {{\rm const}}
\def \V {v}

\def \bv {v^*}
\def \vv {{\rm v}}
\def \LL {{\mathcal L}}
\newcommand{\PV}[1]{P_{\!\!_{V_{#1}}}}

\def \bL {\ell}
\def \M {{\mathcal M}}
\def \N {{\mathcal N}}
\def \S {{\rm S}}
\def \vn {\vec n}
\def \tl {\td \l}
\def \td {\tilde}
\def \Prod {\Pi}
\def \O {{\mathcal O}}
\def \Q {{\rm  Q}}
\def \D {\Delta}
\def \N {{\mathcal N}}
\def\tN{{\tilde N}}

\def \m {\mu}
\def \vs {\vec \s}
\def \ie {i.e.}

\def \cD {{\cal D}}

\def  \le  {\l_{\rm eff}}

\def \rS {{\rm S}}
\def\as{{\a}}
\newcommand{\bra}[1]{\mbox{$\langle #1 |$}}
\newcommand{\ket}[1]{\mbox{$| #1 \rangle$}}

\newcommand{\auth}{AUTHORS}

\def\thb{\bar{\theta}}
\def\Thb{\bar{\Theta}}
\def\barp{\bar{p}}
\def\barq{\bar{q}}
\def\barc{\bar{c}}
\def\bard{\bar{d}}
\def\e{\epsilon}

\def \bi{\bibitem}
\def \la {\label}

\def \l {\lambda}
\def\foot{\footnote}
\def \tl  {{\tilde \l}}
\def \sql {{\sqrt \l}}
\def \adss {$AdS_5 \times S^5$\ }
\newcommand{\rf}[1]{(\ref{#1})}
\def \ov {\over}

\def\th{\theta}
\def\Th{\Theta}
\def\vth{\vartheta}

\def\vth{\vartheta}

\def\ra{\rightarrow}
\def\N{{\cal N}}
\def\F{{\cal F}}
\def\cc{\circ}
\def\eqv{\equiv}

\def\ni{\noindent}
\def \ha{{1\ov 2}}
\def \bw {{\rm w}}

\def\r{{\rm r}}

\def \cT {{\cal T}}
\def \no {\nonumber}

\def \J {\mathcal{J}}
\def \del {\partial}

\def \bps {{\bar \psi}}
\def \sqbl {\sqrt{\bar \lambda}}

\def\dF{\dot{F}}
\def\dG{\dot{G}}
\def\df{\dot{f}}
\def \E {{\cal E}}

\def \S {{\cal S}}
\def \J {{\cal J}}

\def\ms{\mathcal{S}}
\def\mj{\mathcal{J}}
\def\soj{\fr{\ms}{\mj}}
\def \R {{\bf R}}
\def \om {\omega}
\def \tH {\widetilde H}
\def \bE {\bar E}
\def \x {{\cal X}}

\def \hV {{\hat V}}

 \def \bb {\bar \beta}
\def \W {{\cal E}}

\def \bi{\bibitem}
\def \la {\label}

\def \l {\lambda}
\def\foot{\footnote}
\def \tl  {{\tilde \l}}
\def \sql {{\sqrt \l}}
\def \sqtl {{\sqrt {\tilde \l}}}

\def \HH {{\rm E}}
\def \cS {{\cal S}}
\def \cL {{\cal L}}

\def \adss {$AdS_5 \times S^5$\ }

\def \D {\Delta}
\def \thet {\theta}
 \def \t {\tau}
 \def \p {\phi}
 \def \r {\rho}
 \def \rN {{\rm N}}
 \def\tw{{\tilde w}}
 \def\hJ{{J}}
 \def\hw{{w}}
 \def\hl{{\lambda}}
 \def\hth{{\theta}}
 \def\NN{{\cal N}}
 \def \bv {{ \bar w}}
\def \vn {{\vec n}}
\newcommand{\sfrac}[2]{{\textstyle\frac{#1}{#2}}}
\def \bl {{ \bar \lambda}}
\def \bp {{\bar p}}
\def \bu {{\bar u}}
\def \sha {\sfrac{1}{2}}
\def \w {\omega}
\def \ov {\over}
\def \vl { \vec \ell}
\def \varpi {{\rm w}}
\def \OO {{\cal O}}
\def \bG {\bar \G}

\def \c {\gamma}

\def \ss {{\rm s}}

\def \ve {\varepsilon}
\def \pa{\partial}
\def \I {{\cal I}}
\def \LL {{\cal L}}
\def \ep {\epsilon}
\def \R {{\rm R}}
\def \tilt {{\tilde t}}

\def\pic #1#2{\hbox{\lower#1pt\hbox{~\mbox{\epsfxsize=20truemm \epsffile{#2}}}}}

\def\pic #1#2#3{\hbox{\lower#1pt\hbox{~\mbox{\includegraphics[scale=#3]{#2}}}}}

\def \bss {\bigskip}

\def\tx{{\tilde x}}
\def\ty{{\tilde y}}
\def\tf{{\tilde f}}
\def\tz{{\tilde z}}
\def\tpo{{\tilde \phi}_1}
\def\tpt{{\tilde \phi}_2}
\def\hx{{\hat x}}
\def\hz{{\hat z}}

\def\IR{\mathbb{R}}
\def\IC{\mathbb{C}}
\def\IZ{\mathbb{Z}}
\def\IP{\mathbb{P}}

\def\be{\begin{eqnarray}}
\def\ee{\end{eqnarray}}

\def\id{\protect{{1 \kern-.28em {\rm l}}}}

\def\bfsigma{{\boldsymbol{\sigma}}}

\def\lin{{-\!\!\!-\!\!\!\!-\!\!\!\!-\!\!\!-\!\!\!-}}

\def\pint{{-\!\!\!\!\!\!\int}}

\def \bt {\bar\theta}
\def \te {\theta}

\def \cc {{\rm f}}
\def \d {\delta}

\def \cL {{\cal L}}
\def \S  {{\rm S}}
\def \pp {{q}}
\def \vt {\vartheta}
\def \mm {{\cal  \ell}}
\def \Z {{\cal Z}}
\def \pa {\partial}
\def \C {{\cal C}}
\def \be {\bea}
\def \ee {\eea}
\def \c {\gamma}  \def \d {\delta}

\def \DD {{\rm D}}
\def \chii {\varepsilon}
\def \th {\theta}

 \def \t {\tau}
 
\def \beq {\be}
\def \eeq {\ee}
\def \beqa {\bea}
\def \eeqa {\eea}

\def\tx{{\tilde x}}
\def\tz{{\tilde z}}
\def\hx{{\hat x}}
\def\hz{{\hat z}}

\def\fx{~{\tilde x}}
\def\fy{~{\tilde y}}
\def\fpo{~{{\tilde \varphi}_1}}
\def\fpt{~{{\tilde \varphi}_2}}
\def\ff{~{{\tilde \phi}}}

\def\Tr{{\rm Tr}}
\def \cA {{\cal A}}

\begin{document}
\overfullrule=0pt
\parskip=2pt
\parindent=12pt
\headheight=0in \headsep=0in \topmargin=0in \oddsidemargin=0in

\vspace{ -3cm}
\vspace{-1cm}

\rightline{Imperial-TP-AT-2007-2}

\begin{center}
\vspace{0.1cm}
{\Large\bf
Strong-coupling expansion  \\ 
\vspace{0.15cm} 
of cusp anomaly  and    gluon amplitudes\\
\vspace{0.2cm}
from quantum open strings in $AdS_5 \times S^5$
\vspace{0.3cm}
   }

 \vspace{.2cm} {M. Kruczenski$^{a,}$\footnote{markru@purdue.edu},
  R. Roiban$^{b,}$\footnote{radu@phys.psu.edu},
  A. Tirziu$^{c,}$\footnote{tirziu@mps.ohio-state.edu}
 and A.A.
 Tseytlin$^{d,}$\footnote{Also at
 Lebedev  Institute, Moscow.
  tseytlin@imperial.ac.uk
 }}\\
 \vskip 0.03cm

{\em 
$^{a}$Department of Physics, Purdue University,\\
W. Lafayette, IN 47907-2036, USA\\
$^{b}$Department of Physics, The Pennsylvania  State University,\\
University Park, PA 16802 , USA\\
$^{c}$Department of Physics, The Ohio State University,\\
Columbus, OH 43210, USA\\
\vskip 0.08cm $^{d}$  Blackett Laboratory, Imperial College,
London SW7 2AZ, U.K. }

\end{center}

 \begin{abstract}
An important ``observable'' of planar  ${\cal N}=4$  SYM 
 theory is  the  scaling function   $f(\l)$  that appears 
 in the anomalous dimension of large spin  twist 2 operators 
 and also  in the cusp anomaly of light-like Wilson loop.
 The non-trivial relation    between  the anomalous dimension and the Wilson loop 
 interpretations of $f(\l)$ is   well-understood 
 on the perturbative  gauge theory side of the AdS/CFT duality. 
 In the first part of this paper we present  the  dual string-theory counterpart 
 of this relation, i.e. the equivalence  between the closed-string and the open-string 
 origins of $f(\l)$. We argue that the coefficient of the  
 $\log S$ term in the energy of the 
 closed string with large spin $S$  in $AdS_5$ should
  be equal to the  coefficient  in the logarithm of expectation 
   value of the null cusp Wilson loop,
  to all orders  in $ \l^{-1/2}$ expansion. The reason is that the corresponding 
 minimal surfaces  happen to be related   by  a conformal 
 transformation (and an  analytic continuation). 
 As a check, we explicitly compute the leading 1-loop string sigma model correction 
 to the cusp Wilson loop,    reproducing the  same subleading 
 coefficient in $f(\l)$ as  found earlier in the  spinning closed string case.
 The  same function $f(\l)$   appears also in the resummed 
form of the  4-gluon amplitude  as discussed at weak coupling by Bern, Dixon and  Smirnov
 and  recently  found at the leading order at strong coupling by Alday and Maldacena (AM). 
 Here we attempt to extend  the latter   approach to a subleading order in $\l^{-1/2}$
 by  computing the IR singular part of the 1-loop string correction to the
  corresponding T-dual Wilson loop. We discuss explicitly the 1-cusp case
  and comment on  apparent  problems   with the  dimensional regularization proposal 
  of AM  when directly  applied order by order in strong coupling (string inverse tension)
   expansion. 
\end{abstract}
\newpage

\renewcommand{\theequation}{1.\arabic{equation}}
 \setcounter{equation}{0}

\setcounter{equation}{0} \setcounter{footnote}{0}
\setcounter{section}{0}

\section{Introduction}

As is well known, in perturbative (planar $\cal N$=4) gauge theory 
there are  two alternative routes that lead to the same 
scaling function $f(\l)$: it can be found as a coefficient in the 
anomalous  dimension of gauge-invariant large spin 
 twist two   operator \ci{geo,lip} or as a cusp anomaly of a  light-like Wilson line
\ci{pol,dor,kor,mos}.
One can give a  general proof of the equivalence  between the
 two pictures in perturbative  gauge theory \ci{kor}.

On the dual 
perturbative $AdS_5 \times S^5$   string theory side the
 anomalous dimension of minimal twist operator 
is represented by the 
 energy of a  {\it closed}    string with large spin $S \gg 1$  in $AdS_5$ \ci{gkp}, 
$ 
E = S + f(\l)   \ln S + ...\ ,\ 
 \ f(\l)_{\l \gg 1 } = { \sql \ov \pi} + ...\ .  
$ 
The same result  for  the strong-coupling limit of $f(\l)$ was shown in 
\ci{kru} (see also \ci{mak}) to follow from the {\it open} string picture, 
i.e. from the 
 area of a  surface ending on a  cusp 
formed by two  light-like  Wilson lines 
on the boundary of $AdS_5$.

The definitions of $f(\l)$   in \ci{gkp} and \ci{kru} 
 seem very different and a priori unrelated,
 in contrast to the known 
perturbative gauge theory equivalence of the anomalous dimension of
large spin twist 2 operator
and the Wilson loop 
cusp anomaly function.
In particular, while  it was  possible to compute 
the two  subleading  quantum corrections to $f(\l)$ in the closed 
spinning string picture 
\ci{ft1,ftt,rtt}\foot{The expression for  $a_2$  can be found   in \ci{rtt}.}
\be \la{val}
f(\l)_{_{\l \gg 1 }} = a_0 \sql  + a_1 + {a_2 \ov \sql} + ... \ , 
\ \ \ \ \  \ \ \ \ \  a_0={ 1\ov \pi}, \ \ \ \ \ \ \ 
a_1= - {3 \ov \pi} \ln 2 \ , 
\ee
the direct  computation 
of the quantum string corrections in the  Wilson loop approach  \ci{maldrey,dgo,kru}
appeared to be  harder (for previous  attempts in that direction see  \ci{kt,forst,dgt}).

\bigskip

One of our   aims  below  will be
 to  explain the relation between these  two approaches, making their 
equivalence manifest (to all orders in  the strong-coupling,  
i.e.  $1 \ov \sql$ expansion). 
 In particular, we shall  demonstrate  that  computing the quantum 
open      \adss string 
fluctuations  near the cusp surface of \ci{kru} leads to the same 
1-loop  coefficient  $a_1$  in \rf{val}   as found in the closed-string picture 
in  \ci{ft1}.

\bigskip

The  key  observation \ci{ft1,ftt}
that simplifies dramatically 
the  computation of  quantum string corrections to the 
 closed string  energy in the large spin  limit is that in order
 to compute the coefficient of the
leading $\ln S$ term in $E$  it is enough to consider 
a  ``scaling''    limit  of the full (elliptic function)
solution of 
  \ci{gkp}. In this limit 
the string  is stretched homogeneously along the radial  direction of $AdS_5$
all the way to the boundary,  i.e.  
$\rho = \k \sigma$, with $\k \approx { 1 \ov \pi} \ln { S \ov \sql} \gg 1  $.
One  may also ignore  
the boundary (turning-point)  contributions  since 
they are  subleading in the large ${ S \ov \sql}$ limit.
This scaling-limit  solution  is formally related \ci{ftt,rtt} 
 via an  analytic continuation to
 the  circular  rotating string with two equal $S^5$ angular momenta 
 $J_1=J_2$ \ci{ft2,art} and  an imaginary value of the winding parameter. 
 This reveals a simple ``homogeneous'' nature
 of this string background: the only non-trivial 
  fields are isometric angles 
    which are linear in the world-sheet coordinates 
     $(\sigma, \tau)$, i.e.   their derivatives are constant and 
 so  are the coefficients in the fluctuation
 Lagrangian. This  makes the computation of the quantum corrections 
 rather straightforward. 
   
   \bigskip
   
Remarkably, the cusp Wilson loop solution   also 
admits \ci{kru}    a similar simple   ``scaling''   limit 
 in which the Euclidean  open-string  world  surface 
ends  on  two light-like lines on the boundary.
This limit  (combined with a regularization of the  world-sheet area) 
was  enough to reproduce  \ci{kru}
 the leading term $a_0$  in the strong-coupling expansion \rf{val} 
  of the cusp anomaly
 function. 

\bigskip

As we shall explain below in section 2, the two limiting 
 (``scaling'') solutions  are actually closely related:
they become equivalent 
 upon certain analytic continuation (that is  needed in particular to convert 
the Minkowski world sheet coordinates in the closed  
spinning  string case 
into the Euclidean one in the open string Wilson loop 
case)\foot{A similar analytic continuation (in spin)
was used 
to relate the anomalous dimension of twist 2 operators 
to the Wilson loop cusp anomaly picture on the gauge theory side \ci{kor}  (which 
involved also 
 separating the fields in the twist 2 operator  and inserting a Wilson line).}
  and an  $AdS_5$ isometry, i.e. a conformal  $SO(2,4)$ transformation.
   
That relation implies that all quantum
 string  corrections to the two   partition functions 
   which are computed  by 
 expanding the string action  near the equivalent classical solutions 
 should also agree. 
 The 
  logarithm of  the 
  string partition function  in the conformal gauge 
  determines (after dividing by the time interval) 
   the quantum string  corrections to the closed string energy in the scaling limit 
   \ci{ftt,rtt}. The 
   same open-string  partition function on the disc  leads  (after an appropriate  
   regularization  of the world-sheet area \ci{kru}) to  the cusp anomaly function.
  Given the general nature of the relation  between the two scaling string solutions 
  (and their homogeneous  nature allowing one to ignore boundary effects) 
  the equivalence  of the  quantum corrections to the corresponding string 
  partition functions
   and thus to the $f(\l)$ function 
   should thus  extend to all orders in strong coupling expansion.
   This  provides 
   a proof of the equivalence between the two 
    definitions  of  the scaling  function also  on the dual 
   string-theory side,  which is   another 
   remarkable  manifestation of the AdS/CFT duality.

  We will explicitly check  this general statement in section 2.2   by first  
 showing that the null cusp solution  of \ci{kru} 
 has a hidden
  ``homogeneous'' structure: the Lagrangian for 
 string fluctuations near it has indeed   (after an appropriate field redefinition) 
  constant coefficients   and produces  the same characteristic  spectrum as in 
  the scaling limit of the spinning closed string solution in \ci{ft1,ftt}. This implies 
  that   
the 1-loop correction to the 
 null  cusp Wilson loop  leads to the 
 same $a_1$ coefficient in $f(\l)$  in \rf{val} 
 as found in the spinning closed  string picture.

   \bigskip
   \bigskip
   \bigskip

The cusp anomalous dimension of the Wilson loop (which in gauge theory is determined by a UV
singularity) 
governs also \ci{korr} the IR asymptotics of the gluon amplitudes 
in QCD and in $\N=4$ SYM theory. 
The on-shell gluon amplitudes are IR divergent; in dimensional regularization 
$D=4 - 2 \eps, \ \eps < 0$ the IR poles in $\eps$ 
 exponentiate and the double pole is determined 
in terms of the same $f(\l)$ function \ci{ster}.
 Moreover, $f(\l)$ controls also the finite $\ln^2 { s \ov t}$ part   of the  exponentiated
 form of the  4-gluon amplitude \ci{ABDK,bds} (which is apparently determined by 
  conformal invariance   considerations\foot{This
   is apparently no longer so 
  for $n$-point amplitudes \ci{bht}.}
    relating it 
  the IR singular part \ci{dks}).


The same  exponential expression for the 4-gluon amplitude 
was found also at the leading order in 
strong coupling expansion  using the AdS/CFT  correspondence 
in \ci{am}. 
Symbolically, assuming  some IR cutoff $\mu\to 0 $,  
the IR  singular  part of the 4-gluon amplitude may be written as 
\be
\cA_4 \sim  [ \cA_{div} (\mu, s) \cA_{div} (\mu, t)]^2 \ , \ \ \ \ \ \ 
\cA_{div} (\mu, s)  = {\rm exp} \big[ 
- {\textstyle{1 \ov 8}} f(\l) \ln^2 { \mu^2 \ov |s|} - {\textstyle {1\ov 4}} g(\l)  \ln { \mu^2 \ov |s|}
 \big] \ , \la{dii}
\ee
where  $s= - (k_1+k_2)^2$  and $g(\l)$ is a non-universal 
function depending on a choice of the IR cutoff.
As  was found  in \ci{am} at the classical string order 
  using a special  ``dual'' version of the   dimensional regularization  prescription 
\be f(\l) = { \sql \ov \pi} + {\cal O}(1) \ , \ \ \ \ \ \ \ \ \  \ \ \ \ \ \ 
g(\l) = { \sql \ov 2\pi} (1- \ln 2)  + {\cal O}(1) \ . \la{fgg} 
\ee
Our aim  in the second part of this paper (section 3) 
will be to attempt  to extend the   proposal 
of \ci{am} to the quantum  (1-loop) string level.
 One motivation is to explicitly  check 
that it is  the same universal scaling 
 function  $f(\l)$    that indeed appears in the exponential   form of the 4-gluon 
 amplitude  when computed in strong coupling expansion.
 Another  is to  clarify the   structure of the large $\l$ expansion of the 
second scaling  function $g(\l)$ 
with a hope of understanding  the interpolation 
to the 2-loop \cite{ABDK} and 3-loop weak-coupling result for it in \ci{bds}.

\bigskip


\bigskip
\bigskip
\bigskip

The proposal of \ci{am} was   based on starting  with a ``dimensionally extended'' 
analog of
the near-horizon D3-brane  background to which one should apply T-duality 
transformation along the 
$D=4-2 \eps$ longitudinal directions. 

 The use of the 2d duality transformation 
 or, in the target space language,  of the  T-duality transformation
was one of  the    key   observations  of \ci{am}  that suggested a
relation between
the strong-coupling (semiclassical) computation of the 4-gluon
amplitude  
and 
the computation  of the null cusp Wilson loop. 
Thinking of gluons as represented by open strings (attached to D3-branes)
and considering the 
leading strong-coupling approximation  
one may  expect  (by analogy with high-energy asymptotics
of string scattering amplitudes in flat space    \ci{grom})
 that 
their scattering amplitude should be
dominated by a semiclassical string solution depending on some fixed light-like 4-momenta
 $k^m_i$ (i.e. on the $s$ and $t$ kinematic variables
in the 4-gluon case). 
At a qualitative level, 
since for open strings in flat space the T-duality exchanges the 
Neumann and Dirichlet boundary conditions \ci{dai}, 
 it should 
transform a configuration of strings  with 
free ends 
 (thus having specified  conserved target-space  momenta) 
to the one with the ends   fixed at certain positions. 
As discussed in \ci{am},  in the T-dual picture the 
 open string world surface should then 
 end on a closed  contour  with straight sides 
determined by the light-like momenta $k_i$.\foot{Consider a flat target space  and 
the open-string world sheet as 
a  half-plane ($\tau>0, \  - \infty < \sigma < \infty$). 
Then a  string solution 
$x^m \sim   k^m \ln(\sigma^2+\tau^2)$  that is sourced by an 
 external momentum  term
 (originating  from the vertex operator insertions
 as in \ci{grom})
   is a semi-infinite line representing a
point-like string coming from infinity with momentum $k^m$.
The formal  dual   solution 
  ($\partial_a \td x^m = \epsilon_{ab} \partial_b x^m$)   is 
$ \td x^m  \sim  k^m \arctan{\tau\ov \sigma}$
and  near the origin it is a segment of length $ \pi k^m $ instead of
semi-infinite line.  Formally, 
 the 2d duality is an equivalence in the absence of source terms, 
 but the argument should still  go through for sources that are localized, i.e. are  of 
  delta-function type.}
The resulting world surface is then  related (before an  IR regularization)
 by an $SO(2,4)$  transformation \ci{am} to the  cusp  Wilson loop 
surface found in \ci{kru} and discussed 
below  in section 2. 

This T-duality transformation   appears thus to relate the scattering amplitudes  to the
``momentum space'' Wilson loops  \ci{mig,korr,dks}.
 The latter  were first  discussed 
in a closely related  gauge - string duality  context in \ci{poly}. 

\bs 

In general, given a string sigma model with the metric 
$ds^2= G(z) dx^m dx^m  +  dz^2 + ... $   the corresponding 
classical solution  can be  related  
to a classical solutions   in the T-dual metric 
$d\td s^2= G^{-1}(z) dy^m dy^m  +  dz^2 + ... $ via  $y^m= \td x^m$, \ 
$G(z) \del_a x^m = \ep_{ab} \del^b \td x^m$.\foot{There is an  extra  factor of $i$ 
in front of $\ep_{ab}$   in  the case of the  euclidean 2d  signature.}
 The case  of the standard  $AdS_5$ in Poincare coordinates 
 is special\foot{Here we are  ignoring a  possible  modification
 of \adss due to an  IR regularization \ci{am}.
 }
 since here the  original $ds^2= z^{-2} (  dx^m dx^m  +  dz^2) $
  and the dual $ d\td s^2= z^{2}   dy^m dy^m  +  z^{-2} dz^2 $ 
  metrics  are, in fact, related by 
 the coordinate transformation
  (interchanging the boundary and the  horizon): \  $z \to  z^{-1}$. 
 Here one may say that the  2-d duality  acts on  the same space of all
  classical solutions, 
 provided we combine  it with the coordinate transformation $z \to  z^{-1}$.
 In particular, both the original scattering solution and its Wilson loop counterpart 
 may be viewed, at the classical level, 
 as  solutions  of the same $AdS_5$ 
 sigma model, albeit with different boundary conditions. 
 That  does not mean  that solutions  related by such transformation are 
 equivalent (given, in particular, that they  satisfy different 
 boundary conditions in $(\tau,\sigma)$
 and thus describe different physical situations), 
 but some of their properties are indeed   closely related.\foot{Let us mention also
  that the  above 
 T-duality  effectively inverting the $z$-coordinate 
 inverts also   the notion of the UV and IR  singularities:
 the IR  divergences of the amplitudes are mapped  to UV  divergences 
 of the momentum-space Wilson loops 
 (similar relation was already observed at weak coupling, see 
 \ci{dks} and refs. there). Note also  that  in the dimensional regularization 
 framework of \ci{am}  inverting $z$ effectively corresponds to changing sign
 of $\epsilon = { 4-D \ov 2}$.}

 \bs 
 
In section 3.1   we shall 
first ignore the dimensional regularization aspect
(imposing a formal regularization on the world sheet area only at the  very end)
  and 
assume that  the T-duality relation between the  semiclassical 
 world-sheet describing gluon scattering 
  and  the light-like cusped  Wilson loop suggested in \ci{am} extends 
 beyond the classical  level to the full quantum   world-sheet  theory 
 as defined by the \adss  superstring  action  \ci{mt}.
 The  open string scattering solution in the original near-horizon D3-brane 
 background, i.e. in \adss supported by the 5-form flux   should thus be related 
 to the  light-like cusp Wilson loop surface in the  
 T-dual background, i.e. in the  near-core 
 region of the smeared D-instanton solution. 
 The latter  has  the same \adss metric but is  supported  by a  
 dilaton and a  RR 1-form   background. As a result, 
 while the bosonic sigma model part of the
  associated superstring action is the same,  
 the fermionic part is formally different. 
 A simple form of the corresponding superstring action  was found 
 in \ci{kt}
 by applying the 2d duality to the Poincare patch $x^m$ coordinates in the action of \ci{mt}
 (written  in a particular ``Killing-spinor''  $\kappa$-symmetry gauge).
 The transformation $x^m \to \td x^m\equiv y^m$  may be viewed as a quantum 
  change of variables 
 in the string partition function, so the two partition functions 
 (when properly defined to account for the boundary conditions) 
 should  actually agree. 
 Computing the 1-loop partition function in the T-dual  geometry 
 we shall  indeed   find the same  result 
 for the $a_1$ coefficient in $f(\l)$.

 \bs

In section 3.2  we shall address the issue of  how to extend the dimensional 
regularization prescription of \ci{am} to the quantum string level. 
We will  show how one can redo the  1-loop computation
for the  1-cusp  Wilson loop  using the  $D=4-2 \eps$ prescription 
of \ci{am}. We  will find that,   contrary  to what happened at the classical string 
level \ci{am},
keeping $\eps$ finite does not  provide the required  regularization of the result. 

\bs 
In section 3.3 we  shall  discuss several   problems 
with  the ``quantum string'' version of the  dimensional regularization proposal 
 of \ci{am},  suggesting that it  may  not apply
order by order in the inverse string tension ($1 \ov \sql$) expansion. 

\bs


The  Wilson loop  directly related to the 4-gluon scattering amplitude
is the 4-cusp   Wilson loop   which, in the absence of the regularization, i.e. in $D=4$, 
can be found  \ci{am}  by applying a conformal transformation to the 1-cusp \ci{kru}
 solution.  In the  regularized $D=4-2 \eps$  case finding this solution 
 appears to be complicated  (one can no longer use the conformal transformation
 trick). In  Appendix  A we describe the construction of the 
leading  order $\eps$ term in this  ``regularized'' solution
and comment on the structure of the corresponding small fluctuation Lagrangian.

In appendix B we   attempt to resolve the   problems discussed in section 3.3
by suggesting a  modification  of   
 the dimensional  regularization prescription of \ci{am}
 that  may apply order by order in string perturbative expansion.


\section{Strong coupling expansion of the  scaling function:\\
 equivalence of spinning closed string and  null  cusp Wilson loop pictures}

\renewcommand{\theequation}{2.\arabic{equation}}
 \setcounter{equation}{0}
   
   \bigskip 
Below  we shall use the following notation for   coordinates in $AdS_5$ (we shall often 
set its radius to
1). 
The  global  coordinates $(\r,t,\p,\th_1,\th_2)$
\begin{equation}\la{glo}
ds^2= d\rho^2 -\cosh^2 \rho\ dt^2 + \sinh^2 \rho\ (d \phi^2+ \cos^2
\phi\ d \theta_1^2+\sin^2 \phi\ d \theta_2^2) 
\end{equation}
are related to the embedding coordinates $X_M$  $(M=0,...,5)$ on which $SO(2,4)$ is
 acting linearly 
  by 
\be \la{em}
X_0 + i X_5  = \cosh \r \ e^{i t} \ , \ \ \ 
X_1 + i X_2  = \sinh \r \ \cos \p \  e^{i \theta_1 } \ , \ \ \ 
  X_3 + i X_4  = \sinh \r \ \sin \p \  e^{i \theta_2 }  , \ee
  \be\la{xx}
ds^2 =dX^M dX_M  \ , \ \ \ \ \ \ \ \ \ \ 
 X^M X_M  \equiv  -X_0^2-X_5^2+X_1^2+X_2^2+X_3^2+X_4^2=-1 \ .  \ee 
In the Poincare coordinates with the boundary at $z=0$  one has 
  \be \la{poi}
  ds^2 = {1 \ov z^2} ( dx^m dx_m  + dz^2)  \ , \ \ \ \ \ \ \ \ \ \ \ \ 
   x^m x_m  \equiv   - x_0^2 + x_i^2  \ ,\ \ \  \ \ i=1,2,3 \ .  \ee 
 The relation to the 
embedding   coordinates is
  \be \la{reg} 
  X_0 = {x_0 \ov z}, \ \ \ \  
  X_i = {x_i \ov z}, \ \ \ \ 
  X_4= { 1 \ov 2 z} ( -1 + z^2 +  x^m x_m ) \ , \ \ \ \ \
  X_5= { 1 \ov 2 z} ( 1 + z^2 +  x^m x_m) \ . 
  \ee

\subsection{Scaling limits of the spinning closed string  and cusp Wilson loop minimal surfaces 
and their equivalence}

Let us start with recalling  the scaling limit \ci{ft1,ftt}  of the spinning closed string solution
of \ci{gkp}. Here we shall use the    conformal gauge  with Minkowski 2d  signature, 
$ds^2= - d\tau^2 + d \s^2$.  In the limit 
of large spin  
the spinning closed string   \ci{gkp} written  in global coordinates \rf{glo}
   can be approximated 
by 
\bea
\la{sc}
t=\kappa \tau, \quad\quad  \rho=\kappa
\sigma,\quad \quad  \theta_1=\kappa \tau,\quad \ \quad \phi=\theta_2=0, \ \ \ \ \ \ \ \ \ \
 \k \approx { 1 \ov \pi} \ln { S \ov \sql} \gg 1 \ . 
\eea
Here $\th_1$ is the rotation angle and the string is folded and stretched along $\r$  with $\r_*=
\pm \k \pi$ as turning points.
 Rescaling  the  world-sheet coordinates by 
$\k \to \infty$  one effectively  decompactifies  $\s$, i.e. one 
 may  consider the world sheet as a plane. 
The world-sheet area then scales as $\k^2$ and the classical 
 space-time energy  scales as $\k$, i.e. as $\ln S$. The same is  true  to all orders  in 
 quantum $\a' \sim {1 \ov \sql }$ expansion  since the solution  happens to be homogeneous:
 the fluctuation Lagrangian has constant  coefficients. Thus 
 the quantum fluctuation problem is translationally invariant and 
  the quantum effective action or 
 $\ln Z$ is proportional to the area $\sim \k^2$ to all orders in $ {1 \ov \sql }$  \ci{rtt}.
 
 Writing \rf{sc} in the embedding coordinates we get   
(after rescaling $\t,\s$ by  $\k$ and thus assuming that they take 
values in the infinite interval in the limit $\k \to \infty$)
\bea
&& X_0=\cosh \sigma \cos \tau, \quad  X_5= \cosh \sigma \sin\tau  ,\quad \no
 \\ && X_1=\sinh \sigma \cos \tau   , \quad X_2= \sinh \sigma  \sin \tau, \quad \quad X_3=X_4=0 \ , 
  \la{emb}  \\ 
&& X_0 X_2 = X_5 X_1 \la{mb} \ . 
\eea
The  solution \rf{emb} 
thus belongs  to  the same class of  homogeneous string solutions as the rigid 
circular string  found   in \cite{ft2,art}. In fact, it is 
 formally equivalent, upon an analytic continuation 
and re-interpretation  of the parameters, 
  to the background representing the  circular rotating
 string in $S^5$ with two equal angular momenta \ci{ftt}.

\bigskip

Let us now   find  a   counterpart of this solution 
with {\it euclidean}
  2d world sheet. If we set $\t \to -i \t$ then $X_5$ and $X_2$ 
in \rf{emb} become imaginary, effectively exchanging places  in \rf{xx}
while still preserving 
the $(--++++)$ signature of the 6d metric of the embedding space.  
%
Then we can  get a real $AdS_5$ string solution 
with  euclidean  world sheet by simply renaming the coordinates: 
\begin{equation}
X_0'=X_0, \quad X_1'=X_1, \quad X_5'= -i X_2, \quad X_2'=-i X_5,
\quad X_3'=X_3, \quad   X_4'=X_4 \ , 
\end{equation}
where 
$
-  X_0^{'2} -X_5^{'2}+  X_1^{'2}+ X_2^{'2}+ X_3^{'2}+ X_4^{'2}=-1, 
$
i.e. 
\bea  && X_0'= \cosh \sigma \cosh \tau, \quad X_5'= \sinh
\sigma \sinh \tau, \no \\
&& X_1'= \sinh \sigma \cosh
\tau, \quad   X_2'= \cosh \sigma \sinh \tau, \quad  \quad    X_3'=X_4'=0  \ ,  \label{fo} \\
&& X_0' X_5'=X_1' X_2' \la{ide} \ . 
\eea
As we shall  see   below, this euclidean world sheet counterpart 
of the scaling limit of the spinning closed string solution is directly 
related to the null cusp Wilson line solution of \ci{kru}. 
Note that written in the Poincare coordinates  \rf{reg} the solution 
\rf{fo} becomes
\bea 
z'&=&   (\sinh \tau \sinh \sigma )^{-1} \ , \ \ \ \ \ \  \ \ \ \
x'_0= \coth \tau  \coth \sigma 
  \ ,\cr  
x'_1&=& \coth \tau  \ ,\ \ \ \ \ \ \ \  
x'_2= \coth \s  \ , \ \ \ \  \ \ \ \
    x_3=0 \ . \la{fc}
     \ee

\bigskip

\bigskip
Next, let  us review the cusp   solution found in \cite{kru}
in the Poincare coordinates \rf{poi}, i.e. 
\begin{equation}\la{ki}
ds^2=\frac{1}{z^2}(dz^2- d u^2+ u^2 d \xi^2+dx_2^2+dx_3^2) \  ,\ \ \ \ \ \ \ \
x_0= u \cosh \xi, \quad x_1= u  \sinh \xi \ . 
\end{equation}
The limiting  minimal surface ending on two light-like lines at the boundary $z=0$ found in 
\ci{kru} in the static gauge (i.e. with $(u, \xi)$ as world-sheet directions) 
  is 
\begin{equation}\la{spei}
z=\sqrt{2}  u  = \sqrt{ 2( x_0^2 - x_1^2) }  \  ,  \ \ \ \ \  \ \ \   x_2=x_3=0 \ . 
\end{equation}
The corresponding induced 2d metric  has euclidean  signature: \  
  $ds^2=\frac{1}{2}(\frac{d u^2}{u^2}+d \xi^2)$.
To regularize the area of this surface one may assume  that 
 $ \ell < u < L, \   - { \g \ov 2}  < \xi  <  { \g \ov 2}$
where $L, \g \to   \infty, \ \ell \to 0$ 
\ci{kru}.  Then the coefficient in the   area  representing 
the strong-coupling limit of the cusp  anomaly
  reproduces  \ci{kru} 
the same  value of the leading strong-coupling coefficient $a_0$
 in the scaling function \rf{sc} (see also \ci{mak}). 

The same solution in the conformal gauge ($ds^2 = d \t^2 + d \s^2$)  is 
\be \la{spe}
z=\sqrt{2}  u  \ , \ \ \ \ \ \ \ \ \ 
u= e^{ \sqrt 2  \tau}\ , \ \ \ \ \ \  \ \ \ \xi= \sqrt 2 \sigma \  .  \ee
Looking for a more   general solution in the {\it conformal gauge} that  satisfies 
the limiting $z=\sqrt{2}  u$ condition  one finds that it 
is the same as \rf{spe}  
up to  an $SO(2)$ 
rotation in the $(\tau,\sigma)$ plane:\foot{One may also consider a general  2d 
conformal transformation on $(\tau,\s)$;  the resulting 
value of the classical string action (area) will 
be formally  the same before one introduces a regularization.}
\be \la{spej} z=\sqrt{2}  u \ , \ \ \ \ \ \ \ \ 
u= e^{ \a \tau - \b \sigma}\ , \ \ \ \ \ \ \xi= \a \sigma  + \b \tau\ , \ \ \ \ \  \ \ \ 
\a^2 + \b^2 =2 \  ,  \ee
\be \la{kl}
x_0=  e^{ \a \tau - \b \sigma}  \cosh ( \a \sigma  + \b \tau)
  \ , \ \ \ \  \
   \quad x_1= e^{ \a \tau - \b \sigma}  \sinh ( \a \sigma  + \b \tau) \ ,\ \ \ \ \ \ \ \  
    x_2=x_3=0 \ .  \ee
The two natural simple choices are $\a= \sqrt 2, \ \b=0$  in \rf{spe} 
and  $\a=\b=1$ that we will use below.

Using \rf{reg} the 
conformal-gauge solution  can be written in  
embedding  coordinates as 
\bea
&& X_0=\frac{1}{\sqrt{2}}\cosh (\a\sigma+ \b\tau), \quad
X_5=\frac{1}{\sqrt{2}}\cosh (\a \tau-\b\sigma), \no \\
&& 
X_1=\frac{1}{\sqrt{2}}\sinh (\a\sigma+ \b\tau)
, 
\quad  X_4=\frac{1}{\sqrt{2}}\sinh (\a \tau-\b\sigma), \quad \quad
X_2= X_3=0  \ .   \label{cuspi}\eea 
We can get an equivalent form of this   solution
\bea
&& X_0=\frac{1}{\sqrt{2}}\cosh (\a\sigma+ \b\tau), \quad
X_5=\frac{1}{\sqrt{2}}\cosh (\a \tau-\b\sigma), \no \\
&& 
X_1=\frac{1}{\sqrt{2}}\sinh (\a\sigma+ \b\tau)
, 
\quad  X_2=\frac{1}{\sqrt{2}}\sinh (\a \tau-\b\sigma), \quad \quad
X_3= X_4=0  \    \label{cusp}\eea 
 by performing 
a discrete  $SO(2,4)$  transformation 
that interchanges $X_2$ and $X_4$, 
\be \la{lett}
 X_2\to  X_4 \ , \ \ \ \ \ \ \ \ \ 
 X_4 \to  X_2  \ . \ee
 Note that for the solution  \rf{cusp}  \ci{kru}
\begin{equation} \la{gloi} 
X_0^2-X_1^2=X_5^2- X_2^2=\frac{1}{2} \ . 
\end{equation}
The  transformation \rf{lett}  effectively produces a non-zero 
value of $x_2$:  if we ``project''  \rf{cusp}  back to the Poincare  patch 
using \rf{reg}  we get instead of \rf{kl}
\bea  && z=\sqrt{2} \ e^{ \a \tau - \b \sigma} \ , \ \ \ \ \ \ \ \ \ \ \ \ 
x_0=  e^{ \a \tau - \b \sigma}  \cosh ( \a \sigma  + \b \tau) \ , \no \\
&& x_1= e^{ \a \tau - \b \sigma}  \sinh ( \a \sigma  + \b \tau) \ ,\ \ \ \ \ \ \ \  \ \ \ \ 
    x_2 =  e^{ \a \tau - \b \sigma}  \sinh ( \a \tau  - \b \sigma)      \ , 
  \ \ \          \ \ \   x_3=0 \ .  \la{klg} \eea
It is easy to check   that \rf{cusp}  (or \rf{cuspi})
satisfies the string equations  and  the 
conformal gauge 
constraints (in euclidean  2d metric)   written in the  embedding coordinates:
\bea 
&&\del^a \del_a X_M  - \Lambda   X_M =0 \ , \ \ \ \ \  \ \ X^M X_M =-1  \ , \ \ \ \ \ \ \ 
\Lambda   = \del^a X^M \del_a X_M = 2 \ ,  \la{eqa} \\
&&\del_\t X^M \del_\t X_M - \del_\s X^M \del_\s X_M=0 \ , \ \ \ \ \ \ \ \ 
\del_\t X^M \del_\s X_M =0\ . \la{cons} 
\eea
This is thus a special case of  the constant Lagrange multiplier 
($\Lambda=\const$) solutions
for which $X_M$ satisfies a linear   constant-mass 2d equation  discussed 
in \ci{art}. Such string solutions include rigid circular  rotating strings 
and  are effectively ``homogeneous''.\foot{Note that   \rf{cuspi} cannot describe 
 a regular closed string 
since it is not periodic in $\s$ (unless one considers a scaling limit as discussed 
above in which $\s$ is rescaled by a large parameter)
 but it may be interpreted as an open-string solution with 
an infinite range of $\s$.}

Applying another discrete  $SO(2,4)$ transformation 
one can show explicitly that \rf{cusp}  is  a homogeneous  solution, 
i.e. it can be put into the form where 
only the isometric angles of the  $AdS_5$ metric (i.e. the analogs of 
 $t, \th_1,\th_2$  in the parametrization  \rf{glo})   
are non-zero and   linear in $(\t,\s)$. 
As a result, the  fluctuation Lagrangian will  have constant coefficients 
(after an appropriate choice of the  basis of fluctuation
 fields).  To see this explicitly 
let us group $X_M$  as $
(X_0^2 - X_1^2) + (X_5^2 - X_2^2)   - (X_3^2 + X_4^2 )=1$
 and introduce  new   global  $AdS_5$ 
 coordinates as 
\be X_0\pm X_1 = r_1 e^{\pm  p} , \ \ \ \ \ \  
X_5\pm X_2 = r_2 e^{\pm  q} \ ,\ \ \ \ \ \  
X_3 \pm i X_4 =  r_3 e^{\pm i  h} \ , 
 \ee 
where $r_i$   satisfy  $
r_1^2 + r_2^2 - r_3^2 =1  $, i.e.
 \be
r_1= \cosh r  \cos f, \ \ \
r_2= \cosh r   \sin f, \  \  \  r_3= \sinh r \ . \la{hm} \ee 
The $AdS_5$  
metric written in terms of the  independent coordinates $p,q,\g, r, f$ 
becomes:
\be \la{mee} ds^2 = - \cosh^2 r\ df^2  + dr^2
+  \cosh^2 r\ (\cos^2 f\  dp^2 + \sin^2 f\  dq^2)   + \sinh^2 r\ dh^2  \ . \ee 
Our solution is then 
\be r=0,\ \ \ \ f= { \pi \ov 4},\ \ \ \ p  = \a\sigma + \b\tau, \ \ \ \
q= \a\tau-  \b\sigma \ ,  \ \ \ \ h=0 \ , 
\la{hmm}  \ee   i.e. is    homogeneous.\foot{The metric \rf{mee} 
is of course related to $S^5$   metric  by an analytic continuation 
so there is formally  a similar  homogeneous  $S^5$ solution.}

\bigskip 

\bigskip 
\bigskip

Finally, let us now demonstrate  that the   solution  \rf{cusp}   is, in fact, 
 $SO(2,4)$-equivalent 
to the euclidean world sheet version of the scaling limit of the
spinning closed string solution 
 in (\ref{fo}).  Let us choose $\a=\b=1$ in \rf{cusp} and 
 apply   two discrete  $SO(2)$ rotations in the
$(X_0,X_5)$, and $(X_1,X_2)$ planes (which of course preserve the $(--++++)$ metric) 
\bea
&&
X_0'=\frac{1}{\sqrt{2}}(X_0+X_5), \quad \quad  X_5'=\frac{1}{\sqrt 2}(X_0-X_5), \no \\
&&
X_1'=\frac{1}{\sqrt 2}(X_1+X_2), \quad \quad X_2'=\frac{1}{\sqrt 2}(X_1-X_2) \ .  \la{roa}
\eea
The resulting background is then exactly the same as
 in \rf{fo},\rf{ide}.
 
 \bigskip
 
As was  implicit  in \ci{kru} and recently  discussed in detail in 
\ci{am}
the Poincare-patch solution 
\rf{spe}   interpreted in the global coordinates  \rf{cusp},\rf{gloi}  describes a 
surface ending on a closed line   with 
four and not just one null  cusp. 
The presence of the four  cusps was  made clear 
in \ci{am} 
by applying the transformation \rf{lett}   and 
the same  $SO(2,4)$ transformation as in \rf{roa}. The resulting 
  Poincare patch solution is  similar to the one in 
\rf{fc} (with $X_0 \leftrightarrow X_5$, i.e. $\sinh \leftrightarrow
\cosh, \ 
\coth \leftrightarrow \tanh$)  
and it 
 ends on a rectangular contour 
  with 4 null cusps.\foot{The corresponding 
  open-string solution  ending on a rectangular  contour at
 the boundary 
$z=0$ was interpreted in \ci{am} as being 2d-dual (``T-dual'') 
to the semiclassical  solution describing  the massless open string (gluon) 
4-point scattering amplitude with the Mandelstam variables $s=t$.} 
Taking into account the euclidean continuation and  the  subsequent  rotation
in the $(\tau,\s)$ plane
(going from $\a= \sqrt 2 ,\ \b=0$  case in \rf{spe} to the  $\a=\b=1$ in \rf{cusp},\rf{roa})
one may  relate  the origin of the 4 cusps to the presence  of 
the 4 special  points
 in the spinning closed (and folded) 
string surface: the  two ends where the $\s$-derivatives vanish and the center
where the $\tau$-derivatives 
vanish.\foot{One may also think of
the closed string as a combination of two coinciding open strings 
 with the ends at the two folds; this  
also suggests the presence of 4 cusps in the  joint (euclidean) 
world sheet surface.}

\bigskip\bigskip

To conclude, we have seen  that the scaling (large spin) limit \ci{ft1,ftt} 
of the  spinning closed string solution of \ci{gkp} 
is formally equivalent, upon an  analytic
continuation to the euclidean world sheet 
combined   with a discrete $SO(2,4)$ rotation in $AdS_5$,  
to the global $AdS_5$ version of the   null  cusp   solution of \ci{kru}.
This explains the agreement between the leading strong-coupling expressions 
for the scaling (cusp anomaly) function found respectively in 
\ci{gkp} and in 
 \ci{kru}.

This equivalence, combined with the homogeneous nature of the string background,
 implies 
that the correspondence   between the two pictures -- the spinning closed 
string energy (or the minimal twist   anomalous dimension) 
and the cusp Wilson loop -- extends 
also to the  {\it  quantum} string level. 
The homogeneity of the scaling-limit  string solution 
implying   the $(\tau,\sigma)$ translational 
invariance of the quantum fluctuation Lagrangian   means  
that one can  essentially ignore the difference in the boundary conditions
in the closed  and open string cases.
The contributions  of the quantum string fluctuations to the 
energy of the closed string or to the open string partition function will 
be equivalent    since in this scaling limit 
they are the same at each  interior point  of the string world sheet, 
i.e. they  are  proportional to the (regularized) 
area of the world sheet.

To check this  argument,  
below in section 2.2   we shall explicitly find the spectrum of  quadratic 
fluctuations near the cusp  open string surface \rf{cusp}  and verify that   they  
     indeed 
lead  to the same 
1-loop cusp anomaly coefficient $ a_1 $  in \rf{val}  as 
found \ci{ft1} in the 
spinning closed string case. 
Incidentally, that apparently will be  the first explicit computation of the 1-loop 
string correction to a (non-BPS) Wilson  loop surface  done so far in the literature.


\bigskip

Before turning to the discussion of the quantum fluctuations 
let us mention several possible generalizations 
of the  above discussion. 
One immediate extension  is to consider a more general  cusp Wilson loop 
 to include an angular 
momentum $J$ in $S^5$, in direct  analogy with what was done in \ci{ft1,ftt} 
for the  rotating string (large finite $J$ corresponds to operators with 
 large finite twist). \foot{
 The scaling limit \ci{ftt} of the $(S,J)$ folded closed string solution of \ci{ft1} 
 is the following generalization of \rf{sc}: 
 $t= \kappa \tau, \ \  \rho= \sqrt{\kappa^2 - \nu^2} \sigma, \  \ 
 \theta_1 = \kappa \tau, \  \  \varphi= \nu \tau$
 where $\varphi$ is from $S^5$ and $J= \sqrt \l \nu$. 
 As one can easily verify, the  corresponding generalization of  
 the cusp Wilson line solution in 
 conformal gauge  \rf{spe}  corresponding to the metric 
  \rf{ki} with an extra $d \varphi^2$ term  and having euclidean-signature
   world-sheet metric is:
 $ z = a  u, \ \  u= e^{ b \tau}, \ \ \xi = b \sigma, \ \ \varphi= \nu' \tau$, 
 where 
 $ a= { \sqrt 2 \ b \ov \sqrt{ b^2 + \nu'^2}}$.
 The induced metric is then $ds^2 = { b^2 \ov 2} ( d \tau^2 + d \sigma^2)$
 so one is to set $b= \sqrt 2$ to put it into the canonical form as we did in \rf{spe}
 (then  $ a= { 2 \ov \sqrt{ 2 + \nu'^2}}$).
 The parameter $\nu'$  is related to $\nu$ in the closed string case 
  by $\nu'= i {\nu \ov \k}$ (in  going from Minkowski world sheet solution  for a closed string 
  to euclidean world sheet solution  for the open  string in section 2.1 we 
  rotated $\tau \to -i \tau$; we also rescaled $\tau$ and $\sigma$ by $\k$). 
 Other  Wilson loop solutions with rotation in $S^5$ were considered in \ci{tz}.
 }
 It is of interest also to consider the  generalization of the  spinning closed 
 string to spinning string with $n > 2$ spikes \ci{krus}
 to find its scaling (large spin) limit and  to try to 
 identify then a  related  euclidean  open string world surface.\foot{Below we shall also  discuss a generalization of the 
 above cusp solution to the dimensionally  regularized case, following the suggestion 
 of \ci{am}. It would be interesting to find  explicitly the corresponding 
 form of the spinning closed string solution  (dimensional continuation 
 to $D= 4 -2 \eps$  breaks $SO(2,4)$ symmetry so this is non-trivial). 
 In  contrast to the area of the cusp  the energy of the 
 closed string  should be regular in the $\eps\to 0$ limit.}



\subsection{1-loop string  correction to null 
 cusp Wilson loop expectation value}

Let us now go back to  the conformal-gauge 
 solution \rf{spe} or \rf{cusp} with $\a= \sqrt 2, \ \b=0$ 
and compute the spectrum of small fluctuations near it, demonstrating explicitly 
that it is the same as in the case of the scaling limit of the spinning closed string 
in \ci{ft1,ftt}. That  (together with homogeneity of the solution)  will imply   that 
the 1-loop correction to the cusp anomaly is the same as the 1-loop correction 
 to the coefficient of $\ln S$ in the fast-spinning
string energy. 

Let us start   with bosonic fluctuations. 
It is easiest to consider the  string action  in terms of the embedding coordinates, 
though one can get the
same  fluctuation spectrum  using  the  Poincare coordinates
(that will require  a  non-trivial   choice of the quantum fluctuation
fields -- see appendix A).
The euclidean world-sheet Lagrangian  in the conformal gauge  is
\begin{equation}
L=\frac{1}{2}\partial_a  X^M \partial^a  X_M   +  \frac{1}{2} \Lambda(X^M X_M  +1) \ , 
\end{equation}
with the solution \rf{cusp} satisfying \rf{eqa}.
Introducing the fluctuations
\begin{equation}
X_M  \rightarrow X_M + \tilde{X}_M, \quad \quad  \Lambda\rightarrow
\Lambda+ \tilde{\Lambda} \ , 
\end{equation}
we obtain  the quadratic part of the fluctuation Lagrangian 
with $\tilde X_M$ subject to a linear constraint ($\Lambda =2$) 
\begin{equation}
\tilde{L}_2 = \frac{1}{2} \partial_a \tilde  X^M  \partial^a
\tilde  X_M  + \tilde X^M  \tilde X_M \ , \ \ \ \ \ \ \ 
 \ \ \ \ X^M  \tilde X_M =0 \ . 
\end{equation}
The explicit form  of the latter  constraint  is 
\begin{equation}
\tilde{X}_0 \cosh \sqrt{2}\sigma - \tilde{X}_1 \sinh
\sqrt{2}\sigma + \tilde{X}_5 \cosh \sqrt{2}\tau -\tilde{X}_2 \sinh
\sqrt{2}\tau=0 \ . \la{bb}
\end{equation}
Performing the field redefinition $(\tilde{X}_0, \tilde{X}_1)\to (Z_0,Z_1)$
\begin{equation}
Z_0=\tilde{X}_0 \cosh \sqrt{2}\sigma - \tilde{X}_1 \sinh
\sqrt{2}\sigma, \quad \quad Z_1=-\tilde{X}_0 \sinh \sqrt{2}\sigma +
\tilde{X}_1 \cosh \sqrt{2}\sigma
\end{equation}
and similarly  $(\tilde{X}_5,\tilde{X}_2)\to 
(Z_5,Z_2)$, the  constraint \rf{bb}  takes the form 
\begin{equation}
Z_0+Z_5=0 \ .
\end{equation}
This allows us to eliminate $Z_5$ from the fluctuation Lagrangian  which then becomes
\bea
\tilde{L}_2=&-& \del^a Z_0 \del_a Z_0 +\frac{1}{2}(\del^a Z_1 \del_a Z_1 + \del^a Z_2 \del_a Z_2)
-2
\sqrt{2}( Z_1 \del_\s Z_0 - Z_2 \del_\tau Z_0 ) \no \\
 &+& \frac{1}{2} (\partial^a \tilde{X}_3   \partial_a \tilde{X}_3          +
 \partial^a \tilde{X}_4   \partial_a \tilde{X}_4 +2\tilde{X}_3^2+2\tilde{X}_4^2 )   \ . \la{fluu} 
\eea
This fluctuation Lagrangian is essentially equivalent to the euclidean version of the one
 found in 
\ci{ft1,ftt} for the scaling limit of the spinning closed string solution.
Diagonalizing it we get  two massless modes (whose contribution is compensated by that of the 
two massless  conformal gauge ghosts), 
 one mode with mass $2$ and two modes with 
mass $\sqrt 2$. 
In addition,  there are  five  massless modes from the fluctuations in
$S^5$ directions. 

The quadratic fermionic action \ci{mt}  is given   by\foot{For details 
and notation see, e.g.,  \ci{rtt}. In particular, here $M=0, 1, ..., 9$, \   
${\rm s}^{IJ}=
(1,-1)$, \ $\Gamma_*= i \G_{01234} $, etc.}
\begin{eqnarray}
L_{F2}&=& {i }
(\eta^{ab}\delta^{IJ}
-\epsilon^{ab}{\rm s}^{IJ}) {\bar\theta}^I  e\llap/{}_a
\big[
\delta^{JK} {\cal D}_b
-\frac{i}{2}\epsilon^{JK}\Gamma_*   e\llap/{}_b  \big]\theta^K \ , \la{qaq}
\end{eqnarray}
\be
 e\llap/{}_a =   e^A_ M \del_a x^M  \G_A \  , 
\ \ \ \
\ \ \ \ \
 \   {\cal D}_a=
  \partial_a+\frac{1}{4}\omega_M {}^{AB}\del_a x^M  \Gamma_{AB} \ .
  \la{der}\ee
It simplifies  in the $\th^1=\th^2$  $\kappa$-symmetry gauge used also in \ci{ft1,rtt}. 
It is most straightforward  to find the fermionic spectrum (after a 
continuation to euclidean world sheet) 
 in the
coordinates \rf{mee} in which the solution takes the explicitly  homogeneous form \rf{hmm}
and thus  all the coefficients in \rf{qaq} are constant. As in \ci{ft1}, 
we find eight  fermionic modes with mass 1. 

As a result, the logarithm of the 1-loop euclidean partition function is 
(here 
we assume that the rescaled  $\tau$ and $\sigma$ coordinates change in the 
infinite limits) 
\be \la{pi}
\G_1 =  - \ln  Z_1  = 
 V_2 \int{ d^2 p \ov (2 \pi)^2}\   {\Z_1} (p^2)
 \ ,
   \ee
 \be \la{lji}
 {\Z_1} (p^2)   = \ha  \bigg[  \ln (p^2+4)  + 2 \ln (p^2+2) + 5 \ln p^2 
 - 8  \ln (p^2+1)\bigg] \ . \ee
Here the volume $V_2$ 
  factorizes since our
background is 
translationally invariant. 
Note that $V_2= \int^\infty_{-\infty}  d \tau d \s = 2 A_2 $, where 
 $A_2$  corresponds to an  area  of a single cusp  
  (i.e. $V_2$  is twice the  area of a half-plane 
world sheet 
as appropriate for the open string
case).

 Computing the integral one finds  as in   \ci{ft1,ftt,rtt}
\be \la{en}
 \G_1  =   { V_2 \ov 4 \pi}  \int^\infty_0   dv  \    \Z_1 (v) =   a_1\  A_2 \    
 \ , \ \ \ \ \ \ \ \ \ \ \ \ 
  a_1= - {3 \ln  2  \ov \pi}  \ . \ee
For comparison, the value of the classical action is ($ \sql = {R^2 \ov \a'}$) 
\be \la{kj}
I= { \sql \ov 2 \pi} \int^\infty_{-\infty}  d \tau d \s  \  \ha \del^a X^M \del_a X_M  
= { \sql \ov 2 \pi}\ V_2 = \sql  a_0\  A_2  \ , \ \ \ \ \ \    a_0= { 1 \ov  \pi} \ . \ee
The area  of the cusp open-string world sheet
 regularized  as in \ci{kru} is 
$A_2= {\g \ov 4} \ln { L\ov \ell} $, 
where $\g\to \infty $ is a boost parameter for the two lines forming a cusp and $L$ and $\ell$ 
are the IR and the  UV cutoffs.\foot{Imposing 
a cutoff   at $z=\varepsilon \to 0$ in \rf{ki} 
implies that $\g_{max}  \sim \ln { L \ov \varepsilon }$ where $L$ is a cutoff on $u$.
Then $A_2 \sim  (\ln { L \ov \varepsilon })^2$ (see also \ci{mak}).}


\renewcommand{\theequation}{3.\arabic{equation}}
 \setcounter{equation}{0}

\section{Strong-coupling corrections to IR singular part \\ of 
gluon scattering amplitude:\\
dual Wilson loop expectation value}

 
 In section 2.2 we have computed the 1-loop string correction to the 1-cusp Wilson loop 
  and   found that it contains  the same 1-loop corrected $f(\l)$ function 
 as appearing in the energy  of the 
spinning closed string
 (and we  argued that this  agreement should hold  also to all orders in  strong
  coupling expansion).
 We  used  the standard  \adss  superstring action 
 with the 5-form coupling in the fermionic part.

 Below    we shall repeat this  computation starting with  a
 T-dual superstring action which appears in the  context of the proposed relation 
 between  gluon scattering  amplitudes and  Wilson loops 
 in  \ci{am}.
 It was suggested  in  \ci{am} that using   T-duality 
 one can relate the expression  for the 
  4-gluon scattering amplitude to  the expectation value of a certain  
  4-cusp Wilson loop. 
 It  was argued that  the world surface relevant for the 
 4-gluon  high energy scattering   can be found   via 2-d duality   by starting 
 with the null cusp solution of \ci{kru},  ``lifting'' it to global $AdS_5$ coordinates, 
 applying a certain $SO(2,4)$ conformal transformation (a discrete one  plus a ``conformal 
 boost'' depending on $s/t$), 
 and then reinterpreting   the result back  in the Poincare coordinates. 
 
 If we first  formally ignore the  issue of IR 
  regularization,\foot{By IR regularization we mean  the one  that regularizes the massless 
  gluon amplitudes. In the T-dual Wilson loop picture it appears as a ``UV'' regularization 
  at {\it small}  values of the coordinate $z$.
  In addition, the area of the  1-cusp solution has also an ``IR''  
  divergence at {\it large} values of coordinates $y_i$. This ``IR'' 
  divergence is automatically absent in the 4-cusp  case, being  effectively cut
   off by the lengths of the sides 
  of the  contour related to  momentum invariants $s$ and $t$  \ci{am}.}
    we can argue that since 
 the superstring partition function is supposed to be invariant under the global 
 symmetry of \adss  
 and, moreover, since  the quantum fluctuations should not distinguish between 
 the global and the Poincare coordinates (a change of coordinates  in \adss is 
 a local field redefinition 
 of the quantum 2d  fields) it should  not matter  which form 
 of the 1-cusp Wilson loop solution  we use  for the 1-loop computation.

 Then assuming  an ``a posteriori''   cutoff on the area of world surface 
   we shall find  in section 3.1  the same expression 
 for the coefficient of the  IR singular part  of the 
 partition function, confirming at strong coupling that $f(\l)$ 
 that appears in \rf{dii}
 is the same universal  cusp anomaly as found in the 
 spinning closed string and the 1-cusp Wilson loop 
 cases. 
 

 The second part of the proposal of \ci{am} 
states that 
 the counterpart of the IR dimensional regularization on the gauge theory 
 side should be 
 to consider the dual string theory defined in
  a ``dimensionally regularized'' 
 version of the 
  \adss background with the analog of the  boundary having $D=4 - 2 \eps$ dimensions. 
   This proposal worked fine at the 
   level of the classical string theory, i.e. the 
   leading order in strong coupling expansion,  reproducing \ci{am} 
   the same structure of the 4-gluon amplitude as conjectured \ci{bds}
from the resummation of the weak coupling perturbation theory.
 
 In   section 3.2  we shall     explore  how  this ``dual''  version of dimensional
 regularization  works at the {\it quantum}  string level. 
 This  ``built-in''  IR cutoff breaks  $SO(2,4)$ invariance and thus 
 distinguishes between the 1-cusp and 4-cusp solutions, with the latter   being relevant for the 
 physical scattering  amplitude case.  Since  it appears to be hard to find 
  the  explicit ``dimensionally regularized'' 
 version of the 4-cusp solution\foot{One could attempt  to use perturbation theory 
 in $\eps$ (see Appendix A), and even the leading-order form of the solution 
  was  apparently enough at the  classical level \ci{am}.  However, it is not clear why the same
  should apply at the 1-loop level. There is also a technical problem 
  that expanding the exact-in-$\eps$ action near a perturbative in $\eps$ solution will 
  lead to  spurious (off-shell) 2d divergences  and $\k$-symmetry gauge dependence.}
  we shall limit ourselves to the case of 
  the $D=4 - 2 \eps$ regularized  version of the  1-cusp solution (which happens 
  to be  
  a simple generalization \ci{am} of the solution of \ci{kru}). 
  
  Our motivation will be  based on  the expectation 
    that the 1-cusp contribution 
  should be closely related  to the leading singularity of the IR
divergent part (the Sudakov form factor) of 
%
%
the  scattering amplitude,
  but we should  admit    that 
  there is no  obvious argument supporting 
   that beyond the classical string  level.\foot{At the tree 
  level it  was argued  in \ci{am} that the single-cusp   contributions determine 
  the $1 \over \eps^2$ part of the area of the 4-cusp surface. Moreover,  
  it was noticed  in  \ci{buu} that with a special choice   of the ``IR'' cutoff on 
  the length of the two sides of the cusp (equal to $2 \pi \sqrt{-s} $ where $s$ is the
   Mandelstam variable)  which  complements the ``UV'' dimensional regularization
   at small $z$ 
one gets exactly  the same  answer for the two leading singular  terms  
$1 \over \eps^2$ and $1 \over \eps $
 in the 1-cusp area as found in  the  area  for the 
full 4-cusp  surface in \ci{am}. It  is not clear, however, why this observation 
should extend  to the quantum string case.}

We shall then redo the same 1-loop computation as in sections 2.2 and 3.1 for the 
null 1-cusp  solution now  in ``dimensionally regularized'' T-dual
geometry. We shall find that,  contrary to what happened
at the string tree level, 
 the ``dimensional regularization'' prescription  no longer  appears 
to provide a natural cutoff of the result at small $z$. Moreover, 
even if we formally 
assume that the 1-loop area factor should be  again regularized
  in exactly the same way as the 
tree-level one,  we will still  not be able 
to  relate the 1-loop coefficient of the  $1 \ov \eps^2$ pole to
the   1-loop coefficient in the scaling  function \rf{val}.
More generally, there appears to be a problem with  the usual 
RG-type relation between the  
coefficient of the  $1 \ov \eps^2$ pole  and the cusp anomaly  
when applied order by order in the inverse string tension expansion. 

\bs 
 
 
 We shall  comment on  these problems in section 3.3. 
 As we shall discuss, it is not a priori clear
  why  considering string theory in the ``dimensionally regularized''
 background   should provide the required  analog of the 
 IR regularization of gauge theory, with the logarithm of the Wilson loop 
 expectation value  scaling as ${q_1 \ov \eps^2} + {q_2 \ov \eps} + q_3$ at each order in 
 the $\a' \sim  {1 \ov \sql}$ expansion.\foot{At the same time,  one can 
 give a general argument 
 that  the  logarithm of the
 relevant IR singular 
 factor in the  perturbative gauge theory amplitude in $D=4-2 \eps$ should
  have this particular  structure 
 order by order in $\l$ \ci{ster,bds}.}
We shall  conclude  that the    ``dimensional regularization'' 
prescription as formulated in
\ci{am}   does  not appear to 
work  (at least in the most naive way) order  by order 
in the strong coupling expansion. One option is that  
  its application 
 may require a resummation of  ${1 \ov \sql}$ expansion;
  alternatively, it may require 
 some modification, for example,  a modified relation between
  the parameters $\eps$  and $\mu$
 on the gauge and the string sides. 
 This important issue  needs further clarification.


\subsection{1-loop correction  in  T-dual picture:  fermionic action}


Starting   with the \adss  superstring   action   \ci{mt}  written in a particular 
$\k$-symmetry gauge and applying 2d duality along 4 isometric  directions 
of $AdS_5$ in Poincare coordinates it was found in \ci{kt} 
that the resulting    action 
takes a remarkably simple  form 
($m=0,1,2,3; \ s=1,...,6, \ \ z^2= z^s z^s$)\foot{Here we  assume conformal gauge and 
thus ignore the dilaton coupling originating from  the 2d duality transformation.
We also use Minkowski signature on the world sheet.} 
\begin{equation}
\tilde{S}=-\frac{\sql}{2\pi }\int d^2 \sigma \bigg[
\frac{1}{2z^2}(\partial^a y^m \partial_a y^m+ \partial^a z^s
\partial_a z^s)
+ 2 i \epsilon^{ab}  \bar{\vartheta}(\partial_a y^m \Gamma_m+ \partial_a z^s \Gamma_s) \partial_b \vartheta\bigg]
 \ . \la{acc} \end{equation}
Here $\vartheta$  is a Majorana-Weyl  
 10d spinor related to the two original fermionic 
coordinates by a certain $y$-dependent rotation and a  $\k$-symmetry gauge choice 
involving $ (\delta^{IJ} \pm \G_{0123} \ep^{IJ})$. 
Note that the  action \rf{acc}  is exactly  quadratic in  fermions.\foot{
The starting point in \ci{kt}  was the  action in a  special   $\k$-symmetry gauge choice 
found in \ci{kar}. 
An equivalent action   also   becoming  quadratic in fermions
(and having the same structure as \rf{acc})
after the T-duality  along the 4 Poincare patch coodinates 
  was found in  another  $\k$-symmetry ``S-gauge'' in 
Appendix C of \ci{mts}.}  
This  action was interpreted in \ci{kt} as describing a fundamental superstring 
propagating in the  background which is 
 T-dual to the near-core  D3-brane one, 
i.e.  in the  near-core smeared D-instanton  \ci{gg} background.

Below we shall first  rederive the fermionic part of \rf{acc}  by starting   with the  general 
form of the quadratic part of the GS action 
in a type IIB superstring background 
 and specifying it to the case  of 
the smeared  D-instanton background. 
We shall then expand that action  near the homogeneous  cusp solution 
\rf{cusp},\rf{hmm} 
 and show that the resulting fermionic spectrum is the same 
as found  above in section 2.3 from the  quadratic part of the original \adss action  \rf{qaq}.
 This is of course expected on the basis that the two actions are related by the 
 2d duality  transformation. 
 The resulting 1-loop partition  function   is then the same 
 as  in \rf{lji}  and thus it  leads to the same 1-loop term in the coefficient
 function  $f(\l)$.\foot{As  a by-product, we  will thus  provide a direct check
 of  the 1-loop finiteness of the action \rf{acc}.}

To quadratic order in fermions the form of the 
GS superstring action is determined by the generalized covariant  derivative that enters 
the  variation of 
the target space gravitino field.
 The fact that we  consider  a  superstring propagating in a background   which is 
 a 
 type II supergravity solution 
guarantees the   $\kappa$-symmetry of the classical   string action  \ci{howe}.
We shall use the normalization in which the bosonic part of the string-frame
type IIB supergravity Lagrangian  is  
\be \la{hio}
{\cal L}_{\rm IIB}=e^{-2 \phi} \big[R + 4 (\partial \phi)^2 - \frac{1}{12}  H_{MNK}^2\big]
           -   \frac{1}{2} F_M^2 -
	    \frac{1}{12}  F_{MNK}^2 -   
	    \frac{1}{4 \times 5!} F_{MNKLP}^2 \ , 
\ee
and the gravitino supersymmetry transformation rule is ($I,J=1,2$) 
\be
\delta \psi_M^I &=&\DD_M^{IJ}\chii^J \ , \cr
\DD_M &=&   (\partial_M + \frac{1}{4} \omega_{M}{}^{AB} \Gamma_{AB})   - \frac{1}{8} s_3  H_{ABM} \Gamma^{AB}
+ \frac{1}{8} e^{\phi} \big[{\rm F}\llap/{}_{(1)} s_0 + {\rm F}\llap/{}_{(3)} s_1
 +   \frac{1}{2}  {\rm F}\llap/{}_{(5)}  s_0 \big] \Gamma_M
\ee
with the $2 \times 2$
 matrices  $s_3=\sigma_3$,  $s_1 = \sigma_1$ , $s_0 = i \sigma_2$
  and 
${\rm F}\llap/{}_{(n)}=\frac{1}{n!}F_{A_1\dots A_n}\Gamma^{A_1\dots A_n}$.
Then the quadratic fermionic  action   is a   generalization of  \rf{qaq}
\be
L_{F2}&=& {i }
(\eta^{ab}\delta^{IJ}
-\epsilon^{ab}{\rm s}^{IJ}) {\bar\theta}^I  e\llap/{}_a    \DD^{IJ}_b \theta^J \ . \la{qq}
\ee
For 1/2 supersymmetric  supergravity backgrounds  one has
 $ \DD^2=0$ (i.e. $[\DD_M, \DD_N]=0$)
which is the condition of integrability of the  Killing spinor equation; 
shifting $\theta$ by a Killing  spinor is then a global fermionic symmetry
 of the GS action.

While in  this subsection we are ultimately interested in the case 
when  we start   with a 
 D3-brane solution with
   $D=4$  longitudinal directions,   
   in  the next section  we will consider, following \ci{am},  the case 
 when $D=4-2 \eps, \ \eps\to 0$,  so
 let us  keep  $D$  arbitrary for generality. 
 The solution  T-dual to a Dp-brane solution (assuming T-duality is formally applied in 
 all $D=p+1$  longitudinal directions)  
 is the  D-instanton \ci{gg} smeared in $D$ directions  
\be
ds^2_{10} &=&H^{1/2}(z) (dy_D^2+dz_{10-D}^2)\ , \cr
e^\phi&=&H \ , \ \ \ \ \ F_{(1)}= d C\ , \ \ \ \ \ \ \  C= i H^{-1} \ ,    \ \ \   \ \ \ 
e^\phi F_{(1)}= - \ i\ d\ln H  \ ,   \cr
\nabla_{10-D}^2H&=&0~~\rightarrow~~H=\frac{c_D R^4 \td \mu^{4-D}}{z^{8-D}} \ , \ \ \ \ \  \ \ \ \ \ 
R^4 = \l \a'^2  \ . 
\label{sol}
\ee
Here the harmonic function $H$ was taken  in the near-core
 limit and $\td \mu$ is an  analog 
  of the gauge theory renormalization  scale 
  in dimensional regularization.\foot{In the 
  notation of \ci{am}, \ 
  $c_D= 2^{4 \eps} \pi^{3 \ep} \G(2 + \ep), \ \td \mu = {\mu \ov \sqrt{4 \pi e^{-\g}}}$
  and $\a'=1$.}
Note that the original RR scalar  $C$ is 
imaginary for the D-instanton solution.\foot{The dual to $F_{(1)}$  form $F_{(9)}$ 
is real in  euclidean signature case. Let us note also that 
one may check that
 the dilatino variation 
($
\delta\lambda=\frac{1}{2}\partial\llap/\phi\,\chii
+\frac{1}{2}e^\phi {\rm F}\llap/{}_{(1)}\,s_0\,\chii+{\rm other~ fields}
$) vanishes 
if $\chii^1=- \chii^2$ and that the same relation annihilates the gravitino variation.}

The corresponding  $\kappa$-invariant quadratic fermion Lagrangian \rf{qq} 
then takes the form  (cf. \rf{qaq}):
\be
L_{F2}=i(\eta^{ab}\delta^{IJ} -  \epsilon^{ab} {\rm s}^{IJ})
{\bar\theta}^I e\llap/{}_a   \left[
(\partial_b + \frac{1}{4} \omega_{b}{}^{AB} \Gamma_{AB})\delta^{JK}
- \frac{1}{8} i  \epsilon^{JK}\,e^{\phi} {\rm F}\llap/{}_{(1)} e\llap/{}_b 
\right]\theta^K \ . \la{jg}
\ee
Using  that the only nonzero components of the connection for the metric in \rf{sol} 
 are
\be\la{oop}
\omega_{M}{}^{Ai}=-\omega_{M}{}^{iA}=\delta_M^A\partial_i \ln H^{1/4} \ , 
~~~~~i=9-D,\dots,9 \ , 
\ee
it is easy  to show    that
\be
L_{F2}&=&i(\eta^{ab}\delta^{IJ}- \epsilon^{ab }{\rm s}^{IJ})\  H^{1/4}
\partial_a x^A \  {\bar\theta}^I\Gamma_A \Big[\big(
\partial_b+\frac{1}{2}\partial_b\ln H^{1/4}\big)\delta^{JK}\cr
&&~~~~~~~~~~~~~~~~+
\frac{1}{2}\partial_b x^B\partial_i \ln H^{1/4}\Gamma^{i}\Gamma_B\ (\delta^{JK}- i \epsilon^{JK})
\Big]\theta^K \ , \la{acv}
\ee
where  $i=9-D, \dots, 9$,  $x^A = (y^m, z^s)$,  and  $A,B=0,\dots, 9$   are flat indicies. 

We can  eliminate the first abelian connection     term  $ \big(
\partial_a+\frac{1}{2}\partial_a\ln H^{1/4}\big)$
 by redefining
\be
\theta^I = H^{-1/8} \Theta^I~~, \la{onn} 
\ee
thus getting 
\bea
L_{F2}&=&i(\eta^{ab}\delta^{IJ}- \epsilon^{ab }{\rm s}^{IJ})\  H^{1/4}
\partial_a x^A \  {\bar\theta}^I\Gamma_A \Big[
\partial_b \ \delta^{JK} \no \\
&& ~~~~~~~~~~~~~ +  
\frac{1}{2}\partial_b x^B\partial_i \ln H^{1/4} \ \Gamma^{i}\Gamma_B\
 (\delta^{JK}- i\epsilon^{JK})
\Big]\theta^K   \ . \la{ko}
\eea
Next,   we note
 that the only difference between this  action  and the  one for the  
 $AdS_5\times S^5$  supported by the 5-form flux 
 is that the composite connection term $
\partial_b x^B \partial_i \ln H^{1/4}\Gamma_{i}\Gamma_B  (\delta^{JK}-i
\epsilon^{JK})$
is larger by a factor of $8-D \over 4$  (i.e. by  $1+ \ha \eps$ if $D=4-2\eps$). 
 In the absence of regularization, i.e. for $\eps=0$
this connection term  can be  eliminated  as in \ci{kt} by a
rotation of fermions by the same matrix that appears in the solution of the Killing spinor 
equation.
The same transformation can be made also for generic $D$ \foot{ 
The spinor matrix $\Lambda$ here is a function of $S^5$ coordinates
which appears in the expression for the Killing spinors on $S^5$.} 
\be
\Theta^I=(\Lambda^{\textstyle{ 8-D\ov 4}}){}^{IJ}\vartheta^J \ . \la{kp}
\ee
Consequently, the action becomes   simple  when written in 
terms of ${\vartheta}^I$. 

We are still to fix the $\kappa$-symmetry   gauge. Ref. \ci{kt}  used in the original
 D3-brane context the gauge 
$(\delta^{IJ}-\Gamma_{0123}\epsilon^{IJ})\vartheta^I=0$ which amounts to setting 
$\vartheta^1=\Gamma_{0123}\vartheta^2$. The equivalent action  is found  by choosing,
as   appropriate in the  D-instanton context ($\Gamma_{11}^2=1$)
\be
\vartheta^1=\Gamma_{11}\vartheta^2~, \ \ \ \ \ \  \ \ {\rm i.e.} \ \  \ \ 
\ \ \ \   \vartheta^1=\vartheta^2\equiv \vartheta  \ , 
\ee
where we used that the type IIB fermions are Majorana-Weyl with the same chirality.\foot{There 
is a subtlety  in the discussion of the  GS action in D-instanton background 
in that we should have  considered the euclidean  target space and thus formally
complexify the fermions (cf. \ci{gg}). 
In particular, the  rotation needed to eliminate the second line in \rf{ko} 
is complex and the natural gauge choice appears to be 
 $\theta_1= i \theta_2$.}
This finally 
 leads to the same fermionic Lagrangian as in \rf{acc} 
\be
L_{F2}=- 2i \epsilon^{ab} \partial_a  x^A {\bar\vartheta} \Gamma_A \partial_b\vartheta~.  
\la{ree}
\ee
The 2d duality derivation  of \rf{acc} 
in \ci{kt}  implies that the superstring action does not actually contain 
any  higher-order fermionic terms.

\bs

Finally, we are ready to return to the problem of computing the  1-loop correction 
in ``T-dual''  string theory  expanded 
near  the null  cusp solution  \rf{spe}
(or \rf{cusp}  or \rf{hmm}). The bosonic part of the action is the same  \adss, so the 
fluctuation spectrum is the same  as in section 2.2.
To 
 find the corresponding 
  fermionic fluctuation 
 spectrum   it is 
  easiest to go back to the ``unrotated'' form of the  Lagrangian \rf{ko}
and fix the $\theta^1=\theta^2$ gauge. 
We should  also  take into account 
that  we are now interested in the  euclidean world-sheet metric,
i.e. we are 
to replace $\eta_{ab} \to \delta_{ab}, \ \ep_{ab} \to i \ep_{ab}$.
 The resulting fermionic action  has then  constant coefficients\foot{Starting with 
 the  form of the action in  \rf{ree} to get the simple 
  fermionic fluctuation Lagrangian 
one would need to ``undo'' the rotation \rf{kp}, i.e. to apply  a transformation 
$U=\cosh \eta + \sinh  \eta  \  \G_0 \G_1 $ 
with $\eta$ being linear in world-sheet coordinates.} 
and one finds 8 fermionic massive modes with mass 1,
  i.e. exactly the same spectrum as in 
\rf{lji}.
Using a ``direct'' IR regularization of the world-sheet area as in  \ci{kru} 
or in section 2.2 
(with $ \ln^2 \mu \sim  \g \ln { L \ov \ell}$)   
 we 
then reproduce the first universal term in the amplitude \rf{dii}
with the same   function $f(\l)$  as in the cusp anomaly.

\subsection{1-loop correction  in   T-dual picture with 
  dimensional regularization}

Let us now try to generalize the  discussion at the classical string level in 
 \ci{am}   and consider the  dimensionally regularized version of the 
  1-loop computation of the T-dual 1-cusp  Wilson loop  expectation value
  described in the previous subsection.
  
 In the original high-energy scattering set-up we should replace the \adss space 
 by the 
 near-core  D-brane solution with  the total number of longitudinal directions 
 being $D=4- 2 \ep$, i.e. having the metric 
 \be \la{ets}
ds^2_{10} =  H^{-1/2}(z)\ dx_D^2+   H^{1/2}(z) ( dz^2 + z^2   d\Omega_{9-D}^2)\ , \ee
which should be 
supported by the corresponding dilaton and  RR field strength.
To consider the number of longitudinal dimensions $D$ to be 
continuous  may appear  a bit 
odd since the  1/2 supersymmetric type IIB 
Dp-brane solutions  exist only for even integer $D$.
 This problem goes away after we
T-dualize  along the $D$  directions $x^m \to y^m$ in order to
 switch to the 
``momentum-space'' Wilson-loop  description   \ci{am}. 
This leads to the  smeared D-instanton  background
already  given  in  \rf{sol}. 
Ignoring the overall string tension   factor  $T_\eps={
 \sqrt{ c_D  \l \td \mu^{2 \ep}} \ov 2 \pi }$
 coming from the numerator of $H$ in 
\rf{sol}\foot{Note that dimensions are balanced in the string action
provided $\l$ is dimensionless and $y$ and $ z$  have the same dimension as $\mu$, i.e.
the  mass dimension (this was the implicit assumption in \ci{am}). 
This  dimension assignment can be reversed by rescaling by  a power of $\a'$.}
we then get   the following dual   metric 
\be \la{its}
 {\td {ds}}^2_{10} =   z^{- \ep} \bigg[   {dy_{4-2\ep}^2+  dz^2  \ov z^2} 
 +     d\Omega_{5+2 \ep}^2\bigg]\ . \ee
 In the  strict $\ep\to 0$  limit we get 
  back to the \adss metric where the $AdS$ part (cf. \rf{poi}) 
  and the internal sphere part  factorize. 
  This factorization will still be true  at the classical level (assuming, 
  as we will, 
  that one is interested in solutions localized in $S^{5+2 \ep}$), 
  but it will 
   no longer hold at the level of the quantum fluctuation. 
  

\bs

Let us start with generalizing  the light-like cusp solution \rf{spe} 
from the case of the  $AdS_5$ metric  \rf{ki} to the case of the 
 ``regularized''  metric  \rf{its}.
 Assuming that the only non-zero coordinates are again 
 $z,u,\xi$ 
(in the notation of \rf{ki}), choosing the static   gauge with 
$(u,\xi)$ as the world-sheet directions   and making the $\xi$-homogeneous
ansatz for $z$, i.e.  
$z= F(u)$, 
  we find  that the string action is proportional to 
$ \int  d\xi du  u \   F^{-2-\ep} \sqrt{F'{}^2-1} $. This 
 leads to the solution  \ci{am}
\be\la{fz}
 z=F(u) =\sqrt{2+\epsilon} \ u  \ ,\ 
\ee
which generalizes  the solution \rf{spei} of \ci{kru}.
The induced metric   is then 
\be\la{mew}
 ds^2_2 = \frac{1+\epsilon}{ (2 + \epsilon)^{1+\ha \epsilon}} \left[ \frac{du^2}{u^{2+\epsilon}}
  + \frac{1}{1+\epsilon}\frac{d\xi^2}{u^\epsilon}\right]   \  .   
\ee
Changing the coordinates, it can then be put into the standard conformal-gauge form, 
i.e. conformal to 
$ d \tau^2 + d\s^2= d \zeta d \bar \zeta$. 
In general, one can show that  given 
any function 
\be h=h(\zeta)\ ,\  \ \ \ \ \ \ \ \zeta=\tau + i \sigma  \ , \la{hih} \ee 
the background 
\be
z = \sqrt{2+\epsilon}\  u \ , \ \ \ \ \ \ \  \ u= e^{-{1 \ov 2}  (h+\bar h )} \ , 
 \ \ \ \ \ \ \ \ 
\xi = - \sqrt{1+\epsilon}\ \frac{h-\bar h }{2i} \   \la{bca}
\ee
is a conformal-gauge solution, 
i.e. it solves the   string equations and the euclidean  version of
the  conformal constraints. 
The simplest choice   which is a direct generalization 
of the $\eps=0$ solution \rf{spe} 
is found if $h$ is linear in $\zeta$,  i.e. $ h= -\sqrt 2\ \zeta $: 
\be   \la{sii}  z = \sqrt{2+\epsilon}\  u \ 
 , \ \ \ \ \ \ \ \ \ \ \ \ \   u  = e^{\sqrt 2   \tau } \ ,  \ \ \ \ \ \ \ \ 
\xi =  \sqrt{2(1+\epsilon)}\ \s  \ . 
     \ee
     Then the induced metric is 
     conformally-flat (becoming flat for $\ep=0$):\foot{The choice of
      $h$  that corresponds to   flat
   induced metric 
 is  
$
h(\zeta) =   \frac{2}{\epsilon} 
\ln\left[1+\frac{\epsilon}{2} 
 {\frac{(2 + \ep)^{(2+\epsilon)/4}}{(1+\epsilon)^{1/2}}} 
\ \zeta \right] . $
}
 \be \la{ku}
     ds^2_2 =
      N_\eps
       e^{- \sqrt 2 \ep \tau}  (  d \tau^2 + d\s^2)  \
      , \ \ \ \ \ \ \ \ \ \ \ \ \ \ 
    N_\eps   = { 2^{-  \ha { \eps}}  (1 + \ep) \ov  (1 + \ha  \ep)^{ 1 + \ha \ep} }  \ .   \ee 
The  value of the classical (euclidean) action is  regularized  by  $\ep < 0$
at $ \tau \to - \infty$  but  is still divergent  in other limits, 
i.e. it depends on a choice of  a region in $\tau,\sigma$ space.
Alternatively, the action 
 can be computed in the static gauge as in  eq.(3.29) in \ci{am}, 
 i.e. we start with 
 $I= T_\eps  N_\eps  \int d \tau  d \sigma \  e^{- \sqrt 2 \ep \tau}$ 
 and rewrite it  in terms of
  \be \la{y}
  y_\pm = u e^{\pm \xi} =
  e^{ \sqrt 2 [ \tau \pm \sqrt{1 + \eps} \s] } \ , \ee
  ending up with 
\be\la{fgh}
I= T_\eps   K_\eps  \int_0^{u_*}   {dy_+ dy_- \ov 
  2 (2 y_+ y_-)^{1+ \ha \ep}} \ , \ \ \ \ \ \ \ 
T_\eps = { \sqrt{ c_D \l  {\td \mu} ^{2 \eps} } \ov 2\pi} \ , \ \ \ \ \ \ 
K_\eps =
{ \sqrt{1 + \eps  } \ov  (1 + \ha \eps)^{ 1 + \ha \eps}         }
\ , \ee
where
one needs  to assume a  cutoff $y_\pm \leq u_*$ at large $y_\pm $
which amounts  to specifying 
 the lengths of the null lines forming the boundary of the cusp 
 surface.\foot{Our notation for  $\tau,\sigma$ differs from  eq. \ci{am} by 
 $\sqrt 2$ factor; $I$ is euclidean  action, i.e. $I=-iS$ in the notation of \ci{am}.
 The 1/2 factor  in the integral over $y_\pm$  comes from  from the Jacobian.
 }   
As was noticed in \ci{buu},   choosing this cutoff as 
$u_*=  2 \pi \sqrt{-s} $ 
one gets (after formally combining the 4  1-cusp contributions) 
  the same
 answer for the two leading singular
terms   as found in  the  area  for the 
full 4-cusp  surface in \ci{am}\foot{As was noted in \ci{am}, to get this 
singular part of area  one actually needs only  the 
$\epsilon=0$ form of the classical solution.
Ignoring the $\eps$-dependence in \rf{sii}, the  restriction of 
$(y_+,y_-)$ to the square 
$[(0,u_*),(0,u_*)]$ can be easily translated into the domain 
in $(\tau,\sigma)$ space using that 
$y_\pm = e^{\sqrt{2} (\tau \pm \sigma)}$.
 Integrating the square root of the determinant 
of the  induced metric \rf{ku} in this region will 
lead to the same expression for the area
as in the $y_\pm$ coordinates in \rf{fgh}. 
}. Then  we get from  \rf{fgh}
\be\la{fghi}
I=
\frac{1}{\epsilon^2} \frac{\sqrt{\lambda}}{2\pi} 
\sqrt{\frac{\mu^{2\epsilon}}{|s|^\epsilon}} 
               + \frac{1}{\epsilon}\frac{\sqrt{\lambda}}{4\pi}
		     (1-\ln 2) \sqrt{\frac{\mu^{2\epsilon}}{|s|^\epsilon}} + 
		     {\cal O}(\epsilon^0)  \ . 
\ee
\bigskip 

Our aim now will be to compute the 1-loop correction to the $\eps$-regularized 
 1-cusp Wilson loop in the conformal gauge by expanding near  \rf{sii}.
 To study the bosonic fluctuations it   is useful first to change  the coordinates 
in \rf{its}   from $(z,u,y)$ to $(\nu,\b,y')$ as follows:
\be \la{cha}
z= \sqrt{2 + \ep} \ e^\nu \ , \ \ \ \ \ \  \ \ \ \ \ u= e^{\nu + \b}\ ,  
\ \ \  \ \ \ \ \ \  y_k= e^{\nu} y'_k  \ 
. \ee
Then the 10-d metric \rf{its}  takes  the form 
\bea
ds^2_{10}
 =&&    e^{-\epsilon\nu} \bigg( (2 + \eps)^{-1-\ha \eps} \  \left[-e^{2\beta} 
\left(d\nu+d\beta\right)^2 + e^{2\beta} d\xi^2 
                    + (2 + \eps)  d\nu^2 +
		   (dy' + d \nu\ y' )^2_{2-2\eps} \right] \cr 
&&		    + \   
 (2 + \eps)^{-\ha \eps}\  d\Omega_{5+2 \ep}^2 \bigg)  \ , 
\la{mewi}
\eea
where $y'_k$  stand for the rest of the  ``longitudinal'' coordinates
of  the metric. 
Since the 1-cusp solution is  localized  in these   coordinates  and also  in 
 the sphere coordinates,
  the corresponding 
parts of the quadratic fluctuation Lagrangian will  contain  two sets of 
$2-2\eps$  ($y_k'$)  and  $5+2 \ep$  ($Y_s$)  decoupled  massive   modes. 

Note that since  for $\eps=0$  the coordinates $\nu$ and $\xi$ are 
isometric angles in this parametrization, this explains why the 
1-cusp solution was  homogeneous.
Indeed,  the   solution \rf{sii}, i.e. 
\be  \la{clq}
\nu = \sqrt 2 \tau,\ \ \ \ \ \ \ \ \   \xi=\sqrt{ 2(1 + \eps)}\ \s  , \ \ \ \ \ \ \ 
\  \ \b=0 \  \ee
 has  $\nu$ and $\xi$ linear in $\tau$ and $\s$.
 
This homogeneity is apparently  broken for  $\eps \not=0$ by  the 
overall $e^{-\epsilon\nu} \sim z^{-\eps} $ factor in the metric \rf{mewi}. 
Remarkably, it  can  be effectively regained 
 at the level of the quadratic fluctuations
if we properly redefine   the fluctuation fields  when 
expanding near the classical solution \rf{clq} 
\beqa
 \nu   &=& \sqrt 2 \tau + e^{\frac{\epsilon}{\sqrt 2 }  \tau }\ \td \nu \ ,
  \ \ \ \ \ \ \ \
 \b =  e^{\frac{\epsilon}{\sqrt 2 }  \tau }\ \td \b \ , 
 \ \ \ \ \ \ 
 \xi   = \sqrt{2+ 2\epsilon} \big( \s  + e^{\frac{\epsilon}{\sqrt 2 }\tau}
 \ \td\xi\big) \ , \cr 
 y'_k  &=&   e^{\frac{\epsilon}{\sqrt 2 }\tau} \td  y_k \ , \  \ \ \ \ \ \ \ \ \ \ 
 Y_s = e^{\frac{\epsilon}{\sqrt 2 }\tau  }  \td  Y_s \ , \ \ \ \ \ \ \ 
 k= 1,...,2-2\eps, \ \ \ s= 1, ...,5+2 \ep  \ .   \la{reel}
\eeqa 
The  euclidean-signature quadratic 
fluctuation Lagrangian in the conformal gauge will then have constant coefficients.
Indeed, 
 it is   given  by  (up to integration  
by parts and trivial rescalings) 
\beqa
 \td L_2  &=&  \ha (1+\epsilon) \partial^a \td  \xi {\partial}_a \td \xi +
  \ha (1+\epsilon) \partial^a \td \nu {\partial}_a \td \nu 
  - \ha \partial^a \td \beta {\partial}_a \td \beta 
  -  \partial^a \td\beta {\partial}_a\td\nu \cr 
   &&  + \  2\sqrt 2 \big[ 
    (1+\ha {\epsilon})\td \beta \partial_\tau  \td\nu -  (1+\epsilon) \td\beta \partial_\s  \td\xi 
    +\frac{\epsilon}{2} (1+\epsilon)\td \nu  \partial_\s \td\xi  \big]
      \cr
   &&  + \ \frac{\epsilon^2}{4}(1+\epsilon)\td \xi^2 +  \frac{\epsilon^2}{4}
   (1+\epsilon)\td \nu^2 - 
    \frac{\epsilon^2}{4}\td\beta^2 
   - \ha {\epsilon} (4 + 3 \epsilon)
    \td\nu\td\beta \cr 
&&  +\ 
   \ha \partial^a  \td y_k  {\partial}_a  \td y_k +    {(1+\ha \epsilon)^2} \td y^2_k  
   + \ha \partial^a\td Y_s {\partial}_a \td Y_s + \frac{\epsilon^2}{4} \td Y^2_s \ .  \la{jl}
\eeqa
 For $\epsilon=0$  this Lagrangian becomes equivalent  to \rf{fluu}
 (with $S^5$ modes added). 
 
  Thus  even  for $\epsilon\not=0$, 
    the 1-cusp solution  is again effectively 
     homogeneous, and, 
as in the discussion of the non-regularized  case in \rf{kj}, 
we should  then find that the 1-loop quantum correction 
 is   again 
 proportional to   (regularized value 
of)  the world-sheet  area.

It is, however,   clear from the structure of the metric \rf{mewi}
that once we go beyond the quadratic fluctuation (string 1-loop)  level
the interaction vertices will contain  powers of the effective $\tau$-dependent coupling 
proportional to the  inverse of the ``running string  tension'', i.e. 
${ 1 \ov \sql} z^\eps \sim { 1 \ov \sql} e^{\eps  \nu} 
={ 1 \ov \sql} e^{ \sqrt 2 \eps  \tau}$.
While this effective string  tension provided a cutoff at small $z$ (large 
 negative $\tau$) 
at the string tree level, i.e. in the world-sheet area, 
this apparently will no longer  be so  at higher orders 
of inverse  tension expansion. 
The same  conclusion is then expected also in the case of the 
 4-cusp solution of \ci{am}. Thus there appears to be 
  a problem with implementation of the 
idea of this  ``dimensional regularization'', at least  
order by order in ${ 1 \ov \sql}$ expansion. 
We shall return to the discussion of  this issue below.

\bs


 Given that  the action \rf{jl} has constant  coefficients 
 it is straightforward to find the  fluctuation spectrum.
 We shall first   assume that $\tau$ and $\sigma$  
 run in the  infinite  range, i.e.  ignore the presence 
of a   cutoff  on    the  world-sheet  coordinates 
that implements an  ``IR'' cutoff in space-time
(we will need to introduce  it at the end  to regularize  the area factor
as in the 1-cusp case in \rf{kj}). 
We shall also assume trivial boundary conditions on the 2d fluctuation fields. 
 Then we can  
  use the standard 2d momentum 
 representation to diagonalize the 
  fluctuation modes.\foot{A clarification is in order. When 
 discussing a spectrum of quadratic  fluctuations 
 of a string sigma model  we may assume that 
  fluctuations of the  coordinates are normalized as 
 $\int d^2 \s  \sqrt{ \bar g}\  \hat x^M \hat x^N  G_{MN}(x) $.
 The redefinition of the  fluctuations 
  made  in \rf{reel} removes  the $G_{MN}(x) $ factor, i.e.
  replaces $\hat x^M$ by the tangent-space vectors $\td x^M$.
 The fiducial 2d metric  $\bar  g_{ab}$  may be either 
  the original independent metric of  Polyakov's action 
  or the induced metric (on the equations of motion   or in the conformal gauge 
  they    differ only by a conformal factor).
    Since  the induced metric  in \rf{ku} is not flat  for $\eps\not=0$
    one may wonder if we are allowed to consider the standard  
   Fourier-mode basis  when computing the spectrum. 
   The answer is yes, provided all the modes (including the fermions 
   and the  conformal gauge ghosts) are normalized using the same fiducial metric:
   its conformal factor dependence should then cancel out since the theory 
   should not have nontrivial Weyl anomaly (see also  \ci{dgt}).}

   
   \bs

  The three mixed modes $\td\xi,\td\nu,\td \beta$   in \rf{jl} 
   lead to  the following contribution   to the   1-loop effective action, i.e. 
to the  1/2 of  the logarithm of the fluctuation 
  determinant (here $p^2= p_1^2 + p^2_2$;   cf. \rf{pi},\rf{lji}) 
  \be \la{deg}
\ha   V_2  \int{ d^2 p \ov (2 \pi)^2}\   \bigg(  \ln \big[p_1^2 + (p_2  + \ha \eps)^2\big]
+  \ln \big[p_1^2 + (p_2  - \ha \eps)^2\big] 
+ \ln \big[p^2 +  4  + 4 \eps + \ha \eps^2\big] \bigg) \ . \ee  
 The contribution of the first two terms here is  equivalent to the 
   contribution of the  two massless modes, 
 i.e. is  the same as in the $\eps=0$ case. Indeed, using polar coordinates 
 in  $p$-space and integrating  over the angle   using 
$\int_0^{2\pi} d\theta  \ln\left(a+b\cos\theta\right) =
2\pi \ln\left(\frac{a}{2}+\frac{1}{2}\sqrt{a^2-b^2}\right)
$  
we find that 
\be  \int{ d^2 p }\   \big(  \ln [p_1^2 + (p_2  + \ha \eps)^2]
+  \ln [p_1^2 + (p_2  - \ha \eps)^2] \big) 
= 2 \int{ d^2 p }\   \ln  p^2 \ . \ee
This is exactly what is required  to cancel the contribution of the two massless 
modes of the 
diffeomorphism ghosts in the conformal gauge. 

\bs

The action needed to find  the fermionic contribution was already discussed in 
subsection
3.1. Starting with the action \rf{ko} 
 where $H$ is now  given by \rf{sol} with
$D= 4 - 2 \eps$, one  should 
plug in  there the solution \rf{sii} and repeat the same computation of the 
fermionic fluctuation frequencies as described at the end of the 
previous subsection. Using the
 $\theta^1=\theta^2$ gauge  and  noticing  that the abelian connection term drops out since 
 $\bar \theta \G^A \theta=0$, it follows that the quadratic fermion
Lagrangian is
\be
{\cal L}=2\sqrt{2}\,i\,u\left[
 {\bar \theta}(\Gamma_0+\sqrt{2+\epsilon}\,\Gamma_1)\partial_0\theta
+\sqrt{1+\epsilon}\,{\bar\theta}\Gamma_2\partial_1\theta
-\sqrt{\textstyle{\ha}(1+\epsilon)(2+\epsilon)}{\bar\theta}\Gamma_{012}\theta\right]
\ee
The overall factor of $u$ can be eliminated by rescaling of $\theta$
by  $u^{1/2}$. The fermion propagator is then proportional to the
inverse of $p^2+\left(1+\ha\eps\right)$ implying that
we  find 8 fermionic  modes   with mass squared equal to  $1 + \ha \eps$.

\bs

Including the contribution of the  ${2-2\eps}$ longitudinal 
 modes $\td y_k$ and $5+2 \eps$  sphere modes $\td Y_s$ in \rf{jl}  
 we  end up with the following expression for the
  1-loop effective action  generalizing the one in  
  \rf{pi},\rf{lji} to the case of $\ep\not=0$ 
 \bea \la{dega}
\G_{1} &=&\ha   V_2  \int{ d^2 p \ov (2 \pi)^2}\  
 \bigg(  \ln \big[p^2 +  4  + 4 \eps + \ha \eps^2\big]
 + (2-2\eps )\ln \big[p^2 +  2 ( 1 + \ha \eps)^2 \big]\cr 
 && \ \ \ \ \ \ \ \ \ \ \ \ \ \ \ \ \ \ \ \ \ \ \ 
         +\  ( 5+2 \eps) \ln \big[p^2 +  \ha  \eps^2 \big]
-  8 \ln \big[p^2 +  1 + \ha  \eps \big]  \bigg) \ . \eea  
As expected, this   expression  is 2d  UV  finite
(the degrees of freedom and mass  sum rules  $\sum_i  n_i=0, \ \  
 \sum_i  n_i M^2_i =0$ are  satisfied)\foot{This
implies also that the result cannot depend on a mass scale or 
the size of the ``box'' in which $\tau$ and $\sigma$ take values, i.e. it can depend
only of ratios of  masses or scales.} 
 for any $\ep$:
the  D-instanton  background  \rf{sol} 
we started with  is  $\kappa$-symmetric
and thus the  GS superstring action  should be 1-loop finite. 

\bs 
The integral  over  the 2-momentum can be computed explicitly and we get
(with $V_2 = 2A_2$ as appropriate for a single cusp case as in  \rf{en}) 
\be 
\G_{1}  ={ 1 \ov 4 \pi} c_1(\ep)\  {  A_2 }  \ ,\la{gaga}  \ee
\bea
  c_1(\ep)  &=&  \int^\infty _0   dv   
 \bigg(  \ln \big[v +  4  + 4 \eps + \ha \eps^2\big]
 + 2(1-\eps )\ln \big[v +  2 ( 1 + \ha \eps)^2 \big]\cr 
 && \ \ \ \ \ \ \ \ \ \ \ \ \ \ \ \ \ \ \ \ \ \ \ 
         +\  ( 5+2 \eps) \ln \big[v +  \ha  \eps^2 \big]
-  8 \ln \big[v +  1 + \ha  \eps \big]  \bigg) \la{ega}\\
&=&
-\frac{1}{2}\left(8+8\epsilon+\epsilon^2\right) \ln\left(8+8\epsilon+\epsilon^2\right) 
     + 2\epsilon\left(2+3\epsilon+\epsilon^2\right)\ln\left(2+\epsilon\right)
     -\epsilon^2 ({5}+2\epsilon)\ln|\epsilon|  . \no
\eea   
 Expanding for $|\ep| \to 0$  gives 
\be \la{expp}
 c_1(\ep) = -12\ln  2 - 4( 1 + 2\ln 2) \ep   + \ha ( -1 + 9 \ln 2) \epsilon^2   
 -  {5 }\epsilon^2\ln |\epsilon|   + {\cal O}(\ep^3)  \ . 
\ee
For $\ep=0$ we reproduce the value of $a_1= { 1 \ov 4 \pi} c_1(0)$ in \rf{en}.
 Note that the $\epsilon^2\ln |\epsilon| $ term here  originates from the 
 $S^5$ modes that were massless in the $D=4$ case.

  The 2-d  area     $A_2$ 
 that  factorizes  here as in \rf{pi} due to the effective 
 homogeneity of the background 
   is  simply  $\ha \int d \tau d \sigma$, i.e. it does  not contain the 
   $ {1 \ov z^{\eps}} \sim e^{-\sqrt 2 \eps \tau}$  (the effective string tension) 
   factor
   that was present in the  classical   expression for  the area \rf{fgh}. 
   Thus we will need to impose both ``UV''  and ``IR'' cutoffs 
   as in the 1-cusp solution \ci{kru} in \rf{kj}.
   Assuming  a similar conclusion will also apply 
   to the full 4-cusp solution this indicates
    that the
   ``dimensional regularization''   prescription of \ci{am}  fails  
   to regularize the 
   1-loop string correction to the logarithm of the dual Wilson loop.\foot{
 It is interesting to note that the fact that in the absence of regularization the
1-cusp  and 4-cusp solutions are related \ci{am}
by the transformation \rf{roa} combined
 with a ``conformal boost'' imply that the calculation
 above applies to the 4-cusp Wilson
 loop as well, just that the regulator
 factor
  being used in the metric is not the one of \cite{am}.
 Indeed, it is not hard to
find the image of $1\ov  z^{\epsilon}$ through the transformations relating the
1- and 4-cusp solutions.
Thus, given the 1-cusp solution for the Lagrangian
${1\ov z^{2+\eps}} (dz^2 + dy^2 )$
one may construct (through the $SO(2,4)$ transformations that work at $\eps=0$)
a 4-cusp solution for the Lagrangian
$ {R^\epsilon  \ov z^2} (dz^2 + dy^2 )$
where $R$  stands for
        $ -(2 \sqrt{2} z)/[1 + z^2 - 2 (y_0 + y_1 - y_2)
           - y_0^2 + y_1^2  + y_2^2 + y_3^2  ]$.
This     observation reiterates the
fact that the right  string theory counterpart of the regularization
needs to be identified for the correct singular part
of the  amplitudes to be reproduced correctly.}


 \bs 
 
 Despite the redefinition \rf{reel}  that eliminated the $1 \ov z^\eps$ factor from the
 fluctuation action  we may  attempt to 
   consider,  at least at  a heuristic level, 
     a possibility that the 
 area factor $A_2$  in \rf{gaga} should still be  computed using the $\eps$-dependent
 induced  metric in \rf{ku}, i.e. should be  taken as in 
  \rf{fgh},\rf{fghi} but
 without  2 times the string tension  factor $ T_\eps$
 (with extra 2 accounting for the definition of the cusp 
 area as in \rf{kj}, cf. also \rf{fghi}).\foot{
 An attempt to    justify this 
 suggestion could  be based on  the fact that  the fluctuation  fields in \rf{jl}  should 
  be normalized using the induced metric  in \rf{ku}. Then the eigenvalue problem  for 
  the fluctuation operators in \rf{jl} should   be defined  with an extra 
  conformal factor coming from the conformally flat metric \rf{ku}, and that may at the end 
  produce the  area factor  defined  with the non-flat induced metric. However, 
  the condition of cancellation of the total  Weyl anomaly means that  
  one should be able to completely  get rid  of the conformal factor dependence 
  (modulo possibly complications with boundary conditions that we are ignoring here).
  Consider, e.g.,  a 1-loop contribution of a 
  massive scalar in curved conformally flat 2d
  space,  i.e.  $\G_1= \ha \ln \det ( - { 1 \ov \sqrt{g}} \del^2 + m^2 ) $ or, up to a 
  conformal anomaly term, $ \ha \ln \det ( -  \del^2 + M^2 )$, $M^2= \sqrt{g}\  m^2$.   Then 
  $\G_1$ scales as $  \int d^2 \s \ M^2 + ... = \int d^2 \s \ \sqrt{g}  m^2 + ...$.
  In our present case $M^2$ is actually constant, so we do not  get  an extra 
  $\sqrt{g}$ factor in the integral in front of $ M^2 \ln { M_1 \ov M_2}$ type terms in the
  effective action.}
  In this case 
 (and assuming again the momentum  ``IR'' cutoff on the length 
 of the  cusp sides, like  $u_* = 2 \pi \sqrt{ |s|}$)  
 we will have from  \rf{fgh} (using tilde to indicate the result of this
  heuristic prescription)
 \be 
 \td A_2 = { 1 \ov 2\ep^2}  { K_\ep  \ov (2 u_*^2)^{\ha \ep}} 
 =  { 1 \ov 2\ep^2}   -  {\ln ( 2 u_*^2)  \ov 4 \ep } + {\cal O}(\ep^0)  \ . \la{ea} \ee
 Then the  product of $c_1$ \rf{expp}  and  \rf{ea} in \rf{gaga}
 appears to contain   a  surprising  $\ln  |\epsilon|$ term.
 This term should presumably   be omitted:   it is 
 an artifact  of the   procedure of  regularizing  separately 
 the area   and the  fluctuation determinant.
  In practice, one should 
  first take $\eps $ to zero  in $c_1$ and then multiply
 it by the divergent area, i.e. 
 the  product ${ 1 \ov \ep^2} \times \ep^2 \ln \ep^2$ 
  should be  set equal to zero in the $\eps \to 0 $ limit.\foot{
   Note in particular that this issue  is unrelated to 
   the expansion of the  cusp solution in  small $\eps$:  the problem  comes from the
   structure of the dimensionally continued  metric in \rf{its}.
   This problem is more likely  
 to be  a further indication of problems with the
dimensional regularization that we discuss below.}
 Under all these assumptions  one  would 
  finish with  the following expression for the singular part of
 the 1-loop   correction 
 \be 
\td \G_{1}=    - { 3\ln 2 \ov \pi }   \  { 1 \ov 2\ep^2}  
    +  \big[  -  { 1   +     2\ln 2 \ov2 \pi} 
    +   { 3 \ov 4 \pi  } \ln 2 \  \ln ( 2 u_*^2)    \big ]\  { 1 \ov
\ep }+   {\cal O}(\ep^0)   
  \ . \la{aga} 
   \ee
 
 


\subsection{Problems with  IR dimensional regularization
 in string inverse tension expansion
}

As we have  pointed out above, 
the structure of the $D=10$ metric \rf{mewi} with ``running''
effective string tension  
implies that the presence of the $ \eps$ dependent factors will
no longer regularize the quantum string expressions in the $z\to 0$ area
beyond the classical level. 
A related  issue is that of 
applicability of the $\a' \sim { 1 \ov \sql}$ expansion here 
since the curvature of the background   is singular near $z=0$
(cf. \ci{am}).

But even assuming that some modified string theory side  version of  dimensional
regularization prescription  will lead to the  ${b_1 \ov \eps^2} + {b_2 \ov
\eps} + b_3$ 
contributions  like \rf{aga} at each order in the ${ 1 \ov \sql}$ expansion, 
we will still face the following problem    in  matching this
strong-coupling expansion 
with  expectations  based on the perturbative gauge-theory relations
for the  
IR dimensionally regularized gluon  amplitudes.

The general expression \ci{ster} for the IR singular part \rf{dii} of the 
on-shell dimensionally regularized 
gluon  amplitude  applied to the planar limit of the ${\cal
N}=4$ SYM theory yields \ci{bds} 
\be \la{mii}
\cA_{div} (\mu, s)  = {\rm exp} \big[ 
- {1 \ov 8\ep^2 } f^{(-2)} (\td \l)  - { {1\ov 4\ep }} g^{(-1)} (\td \l)  
 \big] \ ,  \ \ \ \ \ \ \ \ \ \    
 \td \l \equiv \l { \mu^{2\ep} \ov |s|^\ep}  \ , 
\ee
  where \ci{rge,ster} 
  \be \la{ff} 
  f(\l) = { d^2 \ov d \ln \l ^2}\ f^{(-2)} ( \l)\ , \ \ \ \ \ \ \ \ \ \ \ \ 
  g(\l) = { d \ov d \ln \l} \ g^{(-1)} ( \l)\ , \ee 
and  $f(\lambda)$ is the cusp (soft) anomalous dimension. 
  These relations were  applied   at  leading  strong coupling  order
in \ci{am} giving   
  the expressions for $f$ and $g$  in \rf{fgg}.

One may argue that  the application of these relations 
at strong coupling  is justified because of  the finite radius 
  of convergence of the planar weak coupling perturbation theory.
   The latter suggests  that the equations \rf{ff} should  be  valid
in a finite  
   disk around the origin in complex $\l$ space 
    from  where they may be analytically continued to the strong 
    coupling region. 
It is not clear, however, why $f^{(-2)}(\lambda)$
should have a regular series expansion at large $\lambda$.
It is consequently unclear whether the relations \rf{ff} 
should hold order by order
    in the strong coupling expansion. 
 
  Indeed, there is  an apparent problem with this suggestion    which appears  when we
  consider the first  subleading order in ${ 1 \ov \sql}$ expansion. 
  Given the relation between gluon
  amplitudes and  ``momentum-space'' Wilson loops proposed in \ci{am}  we expect that 
  the  exponent of $A_{div} (\mu, s) $ in \rf{mii} should have standard string  perturbative 
  expansion,  i.e. that  both    $ f^{(-2)}$ and $g^{(-1)}$   should   scale as 
    $ k_0 \sql + k_1   + { k_2 \ov \sql} + ...$   when  expanded at large $\l$.
     On the other hand,  the  same  pattern   of strong coupling expansion 
    is known   to apply to  $f(\l)$ \ci{gkp,ft1,rtt} (and should  presumably 
     be true  also for    $g(\l)$).  
    This  is,  however,  inconsistent 
    with  the relations 
  \rf{ff}.  Indeed,  to reproduce the expansion \rf{val}, i.e.   $f(\l)
  = a_0 \sql + a_1 + {a_2 \ov \sql } + ... $   using \rf{ff} 
   we need to assume that $f^{(-2)} ( \l)$ should have   the following  behavior
    at large $\l$
   \be \la{ssf}
   f^{(-2)} ( \l\gg 1 ) =  4 a_0 \sql  + { 2a_1 }  (\ln \sql)^2  + {4 a_2 \ov \sql} + ...  \  . \ee 
   Here the first term has the expected classical string form \ci{am} 
   but the first subleading term is  $(\ln \sql)^2$ 
   instead of  a constant.\foot{We can of course add also a
   constant  and as well as  $\ln \sql$    term  to $f^{(-2)} ( \l)$
as this  will 
   not  change  the expression for $f(\l)$.} 
   Such  logarithmic term 
    can not   appear as a perturbative   string sigma correction.
    One   may conjecture that 
   it originates from a  resummation of all  higher-order strong coupling
corrections.\foot{We acknowledge  a discussion with J. Maldacena on this issue.} 
   Another possibility is that \rf{mii}  with \rf{ff}  do  not actually apply
   directly to the strong-coupling expansion  as defined by the perturbative 
   string theory. 
   

And vice versa, even assuming that \rf{aga} represents the 1-loop
contributions to $f^{(-2)} ( \l)$ and $g^{(-1)} ( \l)$, the latter are
then constant ($\l$-independent) and thus, according to \rf{ff}, do
not change $f(\l)$ (and $g(\l)$) at all, in contradiction with
\rf{val}.

\bs

One abstract possibility to reconcile the strong-coupling expansions
of $f^{(-2)} ( \l)$ and of $f(\l)$ while preserving their relation in
\rf{ff} could be to redefine $\lambda$ by a constant that eliminates
the constant $a_1$ term in $f(\l)$.  Namely, if we shift the string
tension by a constant $\sql + {a_1\ov a_0} \equiv \sqrt{ \l'}$ and
identify
$\l'$ with
the gauge-theory coupling we would have no 1-loop correction in both
$f(\l)$ and $f^{(-2)}$.  Shifting the string tension by a constant seems
to lead to problems with various other comparisons between the \adss 
string theory and the $\cal N$=4 supersymmetric gauge
theory but in this particular setting it may seem we have little
understanding of how the two couplings should actually be related.

In appendix B we shall present  another  attempt to modify 
 the dimensional  regularization prescription of \ci{am} 
 by relaxing the identification between the $\eps$ parameters 
 on the gauge theory and the string
 theory sides, which may help to resolve the above problem.

  \bs

Regardless of the problems discussed above,  an 
  unwelcome feature of the   string-side version 
of the IR dimensional regularization  suggested in \ci{am} 
is that instead  of providing a  simple modification 
of the known \adss  string action (by analogy   with what is 
done to define the  dimensionally  regularized gauge theory)\foot{That 
would  be the case if one would simply impose a cutoff 
 at small  $z$, by, e.g., $z^{-2} \to (z^2 + \varepsilon^2)^{-1}$
(cf. \ci{mak,abel}). But then one would need to identify the corresponding
 regularization on the gauge theory side. 
Further complications with
such a regularization include difficulties in finding the minimal
surface in the presence of the regulator as well as in 
finding a consistent Green-Schwarz action. } 
 it instructs  us to   start with a 
 dimensionally-modified supergravity background 
 and rederive the superstring action  from scratch. This is  rather  
 daunting task beyond the
 quadratic level in fermions  which casts doubt on a
practical utility of this prescription.
 For example,  extending the above 1-loop
 computations to the 2-loop string theory 
 level would be  quite  complicated.

 \bs

\noindent
 Note Added: 

An interesting  suggestion of how to  reproduce the $(\ln \sql)^2$  term in
\rf{ssf} (and also how to treat higher-loop orders) 
 was suggested   to us by J. Maldacena (private communication).
The effective string  sigma model  coupling that determines  the
higher-loop order corrections (after we rescale  the fluctuation
fields
 as in \rf{reel} to put  the propagator term  into a  canonical form) is  $
  z^\eps  \ov \sql$  which grows  near $z=0$  for $\eps <0$.
    At one loop we  may then  cut off $z$ at a  minimal
  value $z_{\rm o}$ where  $  {z^\eps_{\rm o} \ov \sql}  $ is fixed to  some
  given value $k$, i.e.
   $\ln z_{\rm o} = { 1 \ov \eps} \ln (k  \sql)  $.  In this case the
 area   factor of a single cusp surface
  which we got in \rf{gaga}
 written in the same way as in \rf{fgh}  can be regularized as follows
(ignoring terms of subleading  orders in $\eps$):
 $A_2 = \ha  \int d \tau d \sigma \to
     \ha   \int
     { dy_+ dy_-  \ov 2   (2y_+ y_-)} $, 
where the integral  should be cut off at large $y_\pm$   at $u_*$ and small $y_\pm$  by 
  the above cutoff  at small $z$. Namely,   
$u= \sqrt{ y_+ y_-}  > u_{\rm o}={ 1 \ov \sqrt 2} z_{\rm o}$
implies that one is to integrate over a ``triangle''  instead of a ``square'' 
in the $\ln y_\pm$ plane, 
$\ln y_+ + \ln y_- = 2 \ln u  >   2 \ln u_{\rm o}  $, and that 
gives an extra factor of 2  compared to  \rf{fgh}.
As a result,  $A_2= { 1 \ov 4}  (\ln { u_*\ov u_{\rm o}})^2
= { 1 \ov 4 \eps^2 } (\ln \sql )^2 + ... $. We then have
according to \rf{gaga}  and \rf{expp}\ 
$\G_1= a_1 A_2 + ... =   { a_1 \ov 4 \eps^2 } (\ln \sql )^2 + ... $
which matches  the result  for the contribution of the one of 4 divergent
 factors in the amplitude \rf{dii} that follows from
 \rf{mii},\rf{ssf}.
For higher  orders one may attempt to analytically continue 
in $\eps$, effectively reversing its sign; replacing then 
$1 \ov \sql$  by $z^\eps \ov \sql $ with $z \to  \sqrt{2 y_+ y_-}$
 under the world-sheet integral  at each
loop order then appears to  lead to coefficients in $f^{(-2)}(\l)$  consistent with \rf{ff}. 
It still remains to be seen  if there is a systematic version of
dimensional regularization that may  allow one to
  capture the strong-coupling expansion  not only
for  $f(\l)$ but  also for $g(\l)$.

\section*{Acknowledgments }

We are  grateful to  L.F. Alday, Z. Bern, L. Dixon, S. Frolov,
G. Korchemsky, D. Kosower,  T. McLoughlin and
 especially J. Maldacena 
for useful communications and discussions. 
The work of M.K. was supported by the National Science Foundation
under grant  No. PHY-0653357.
R.R.   acknowledges
the support of the National Science Foundation under grant
PHY-0608114. 
A.T. was supported in part by the DOE grant DE-FG02-91ER40690.
 A.A.T. acknowledges  the support of
the PPARC, INTAS 03-51-6346, EC MRTN-CT-2004-005104   and the RS
Wolfson award. 
\bigskip

\renewcommand{\theequation}{A.\arabic{equation}}
\renewcommand{\thesection}{A}
 \setcounter{equation}{0}
\setcounter{section}{1} \setcounter{subsection}{0}

 \section*{Appendix A:  
Construction of  ${\cal O}(\epsilon)$ correction to the\\ 4-cusp solution}

In this Appendix we describe the construction of the ${\cal
O}(\epsilon)$ correction to the 4-cusp Wilson loop solution 
(found  in original \adss ``unregularized'' form in
\ci{am}) 
and outline the
structure of the quadratic fluctuation Lagrangian.

To set up the stage, consider the general Lagrangian
\be 
L_\epsilon=[h(\Phi)]^\epsilon L_{\epsilon=0}(\Phi)
\ee
 and let 
$\Phi=\Phi_0$ be  a solution of the Lagrangian $L_{\epsilon=0}(\Phi)$. We
are interested in finding the function $\Phi_1$ such that
$\Phi_\epsilon=\Phi_0+\epsilon\Phi_1$ is a solution of the Lagrangian
$L_\epsilon$ to leading and next-to-leading order in the expansion
in  $\epsilon\to 0$. As mentioned previously, it is not a priori clear
that such an expansion is sufficient for testing the conjectures of 
\ci{am} and \ci{bds}; still, it  gives some information on a class of
functions that may appear in the complete solution.

The strategy
is relatively straightforward. First,  we expand
$L_{\eps=0}$ around $\Phi_0$. The inclusion of the additional 
factor $h^\epsilon $ modifies the equations of motion of
fluctuations by potential terms. The solution to these deformed
equations yield the desired correction. 

Indeed, the expansion of $L_{\epsilon=0}(\Phi)$  around the solution
$\Phi=\Phi_0$ has the following structure:
\be
L_{\epsilon=0}(\Phi_0+\epsilon \phi)=L_{\epsilon=0}(\Phi_0)
+\epsilon \partial_a(c^a\phi)
+\epsilon^2\phi\cdot K(\Phi_0)\cdot \phi
+\epsilon^2\partial_a (\phi\cdot V^a(\Phi_0)\cdot \phi)
+{\cal O}(\epsilon^3)
\ee
The first term here  is simply the value of the
Lagrangian on the classical solution. The second and fourth terms are total
derivatives that integrate to zero at the level of the action. 
$K(\Phi_0)$ is the kinetic operator of the small fluctuations
around $\Phi_0$ and, in general, it may be a function of the world sheet
coordinates. 

With this starting point it is easy to understand the structure of the
equations determining $\Phi_1$. Indeed, in the presence of
$h^\epsilon$ the field configuration $\Phi_0$ is no longer a
solution of the equations of motion. Instead, at the level of the
action, the fluctuations $\phi$ exhibit a tadpole which, to leading
order in $\epsilon$ is just
\be\la{tap}
L_{\rm tadpole}=-\epsilon\,\phi\, c^a\partial_a \ln h(\Phi_0)\ . 
\ee
Thus, to summarize, the equations determining $\Phi_1$ are the
equations of motion of the fluctuations around $\Phi_0$ at
$\epsilon=0$ deformed by a potential generated by the tadpole
Lagrangian \rf{tap}.

In our case we have  $h(\Phi)\equiv 1/z$ and the undeformed solution $\Phi_0$
is the 4-cusp solution of \cite{am} in the conformal gauge
\be
z_0&=&\frac{a}{\cosh u_2 \cosh u_1 + b \sinh u_2 \sinh u_1}
\cr
y_{00}&=&\frac{a\sqrt{1 + b^2} \sinh u_2 \sinh u_1}
{\cosh u_2 \cosh u_1 + b \sinh u_2 \sinh u_1}
\cr
y_{01}&=&\frac{a \sinh u_2 \cosh u_1}
{\cosh u_2 \cosh u_1 + b \sinh u_2 \sinh u_1}
\cr
y_{02}&=&\frac{a \cosh u_2 \sinh u_1}
{\cosh u_2 \cosh u_1 + b \sinh u_2 \sinh u_1}
\cr
y_{03}&=&0~~
\label{WL4}
\ee
where  the parameters $a$ and $b$  are related to the
Mandelstam invariants  \ci{am}  and $(u_1,u_2)=(\tau,\sigma)$.

The relation between the 1-cusp  and 4-cusp solutions implies that a
convenient choice of the basis of  fluctuation fields  
\be
z=z_0+\epsilon \delta z~,  \ \ \ \   ~~~~~y_i=y_{0i}+\epsilon \delta y_i
\ee
is given by $(\fy, \fpo, \fpt,\ff, \fx)$   
\be
\delta z&=&\frac{a}{2(\cosh u_2 \cosh u_1 + b \sinh u_2 \sinh u_1)^2}
\Big[
-2\sqrt{1 + b^2} \fy  \cr
&& -(\cosh u_1 \sinh u_2 + b \cosh u_2 \sinh u_1 )(\fpo+ \fpt)     \cr
&& -(\cosh u_2 \sinh u_1 + b \cosh u_1 \sinh u_2 )(\fpo- \fpt)     \cr
 && + 2 (b \cosh u_1 \cosh u_2 + \sinh u_1 \sinh u_2) \ff
		\Big]
\cr
\delta y_0&=&\frac{a}{4(\cosh u_2 \cosh u_1 + b \sinh u_2 \sinh u_1)^2}
\Big[4 (b\cosh u_1 \cosh u_2 - \sinh u_1 \sinh u_2)\fy  \cr
&&+  \sinh(2 u_1)(\fpo  + \fpt ) 
  +  \sinh(2 u_2)(\fpo  - \fpt )\cr
&&-2 \sqrt{1 + b^2}(\cosh(2 u_1) + \cosh(2u_2))\ff  
\Big]
\cr
\delta y_1&=&\frac{a}{2(\cosh u_2 \cosh u_1 + b \sinh u_2 \sinh u_1)^2}
\Big[
-  2\sqrt{1 + b^2} \cosh u_1 \sinh u_2 \fy \cr
&&+ \cosh^2u_1(\fpo+\fpt)-b \sinh^2 u_2(\fpo-\fpt)
\cr
&&- (\sinh (2 u_1) - b \sinh( 2 u_2)) \ff 
 \Big]
\cr
\delta y_2&=&\frac{a}{2(\cosh u_2 \cosh u_1 + b \sinh u_2 \sinh u_1)^2}
\Big[-2\sqrt{1 + b^2} \cosh u_2 \sinh u_1 \fy \cr
&&
 + \cosh^2u_2(\fpo+\fpt)-b \sinh^2 u_1(\fpo-\fpt) \cr
&& - (\sinh (2 u_2) - b \sinh( 2 u_1)) \ff 
 \Big]
\cr
\delta y_3&=&\frac{a~ \fx}{\cosh u_2 \cosh u_1 + b \sinh u_2
\sinh u_1} 
\label{fluct}
\ee
This choice leads to constant coefficients in the operator $K_0$.

It is straightforward (though somewhat tedious) to find the
 equations corrected by the presence of the regulator. Only
the solution obeying the Virasoro constraints is  physically
interesting. Since for $\epsilon=0$ the solution  does obeys the constraints,
the resulting conditions  on the correction are not sensitive to the regulator.
Indeed, it turns out that the vanishing of the 2d stress tensor of the 
fluctuation fields
 requires 
that $(\fy,\fpo,\fpt,\ff,\fx)$ are related by
\be
(\partial_{u_1}-\partial_{u_2})\fpo-(\partial_{u_1}+\partial_{u_2})\fpt&=&0
\cr
(\partial_{u_1}+\partial_{u_2})\fpo+(\partial_{u_1}-\partial_{u_2})\fpt&=&4\ff~~.
\label{Vir}
\ee
The solution of these equations 
 can be  parametrized by a single arbitrary function $G$:
\be
\fpo=(\partial_{u_1}+\partial_{u_2})G\ , 
~~~~~~~~
\fpt=(\partial_{u_1}-\partial_{u_2})G\ , 
~~~~~~~~
(\partial_{u_1}^2+\partial_{u_2}^2)G=2\ff~~.
\label{Virgg}
\ee
Further using the Virasoro constraints to simplify the remaining equations of
motion it is not hard to see that they in fact decouple. The remaining
functions $(\fx,\fy,\ff)$ are  determined by
\be
&&
(\partial_{u_1}^2+\partial_{u_2}^2-4)\ff
-\frac{\sinh u_1\sinh u_2+b\,\cosh u_1\cosh u_2}
{\cosh u_1\cosh u_2+b\,\sinh u_1\sinh u_2}=0\ , 
\cr
&&
(\partial_{u_1}^2+\partial_{u_2}^2-2)\fx=0\ , 
\cr
&&
(\partial_{u_1}^2+\partial_{u_2}^2-2)\fy-\frac{\sqrt{1+b^2}}{\cosh
u_1\cosh u_2+b\,\sinh u_1\sinh u_2}=0 \ . 
\label{eom}
\ee
The solution for $\ff$ is then the source for the equation \rf{Virgg}
determining $G$. 

The $\epsilon=0$ solution already obeys the boundary conditions that
relate its shape to the Mandelstam invariants of the scattering
process. The corrections to the solution must therefore leave these
boundary conditions unchanged. For the particular case of  $s=t$
(i.e. $b=0$) this translates into
\be u_2\rightarrow\pm\infty :  \ \ \ \ \ \ \ \ \ 
\pm 2\ff=\fpo+\fpt~,~~ ~&&
~  ~~~\fy=0
\cr u_1\rightarrow\pm\infty :  \ \ \ \ \ \ \ \ \ 
   \pm 2\ff=\fpo-\fpt~,~~~&&
~  ~~~\fy=0
\ .
\label{bc}
\ee

Once a solution to \rf{Virgg}, \rf{eom} and \rf{bc} is found, it is
then straightforward to find the value of the classical action. Since
the solution correctly captures only the ${\cal O}(\epsilon)$
deformation, the expansion of the classical action should be truncated
to this order. The result reads
\be
L_{\epsilon}(\Phi_\epsilon)=\frac{1}{z_0^\epsilon}\big[2+4\epsilon
\partial_{u_1}\partial_{u_2}G+{\cal O}(\epsilon^2)\big] \ . 
\label{class}
\ee
Note that $\frac{1}{(z_0+\epsilon\delta z)^\epsilon}=\frac{1}{z_0^\epsilon}+{\cal O}(\epsilon^2)$ so the regulator does not contribute to this order except for the overall factor. 


\bs 

With the ${\cal O}(\epsilon)$ correction to the classical solution at
hand, one may now proceed to compute the quadratic fluctuation
Lagrangian. 
Only the terms of the order  $\epsilon^0$ and $\epsilon^1$ are
reliable.  

The idea is to use as much as possible the fact that the expansion of
$L_{\epsilon=0}$  around the solution \rf{WL4} has -- up to total derivatives --  constant
coefficients if the fluctuation fields
are chosen similarly to eq.\rf{fluct}. With this in mind, the
structure of the quadratic fluctuation Lagrangian contains three types
of terms:

1) the quadratic fluctuation part of  $L_{\epsilon=0}$ 
expanded around the $\eps=0$ solution

2) the cubic vertices of $L_{\epsilon=0}$ 
 expanded around the $\eps=0$ solution in which one of the
fields is replaced by the ${\cal O}(\epsilon)$ correction constructed
above

3) the derivative of $L_{\epsilon}$ with respect to $\eps$ evaluated
on the undeformed solution and expanded to quadratic order in
fluctuations around it.

The terms of the first type have constant coefficients; the coordinate
dependence in the terms of the second type arises entirely from from
the ${\cal O}(\epsilon)$ correction to the solution. The position
dependence in third type of terms arises from the regulator as well as
from the total derivative terms that integrate to zero in the absence of
the regulator. 

The calculation of the determinant of this operator is non-trivial  
 due to a complicated position dependence. An
expansion in $\epsilon$ is,  however,  possible. 

\bs

Still, it is important to stress   that it is unclear whether such an 
expansion yields the expected result for the poles in $\epsilon$. First,
the leading order result has the same unfortunate features discussed
in section 3 -- namely,  that it seems not to be regularized by 
non-zero $\eps$. Also, the orders of poles in $ \eps$ 
appearing at higher orders 
strongly depend on the large distance structure of the correction to 
the $\eps=0$ solution. It is possible that the higher order
corrections bring increasingly higher order poles and the
expected pole structure of the IR singular part of the amplitude \ci{bds} 
arises only after all these poles are
resummed. 

To conclude, it  appears  that a solution exact to all orders in $\eps$ is
required for such an approach to have a chance of leading to a finite
contribution to the expectation value of the  4-cusp Wilson loop.

\renewcommand{\theequation}{B.\arabic{equation}}
\renewcommand{\thesection}{B}
 \setcounter{equation}{0}
\setcounter{section}{1} \setcounter{subsection}{0}

 \section*{Appendix B:  
An attempt of modified dimensional \\ 
regularization prescription}

We have seen in section 3  that   a  naive application of the 
regularization scheme suggested in \cite{am} to the single-cusp momentum 
space Wilson line does not appear to  reproduce the structure of the gauge 
theory Sudakov form-factor $A_{div}\equiv ({\cal M}^{[gg\rightarrow 1]})^{1/2} $
beyond the leading strong-coupling order. 
It turns out,  however,  that it is possible 
to suggest  a formal prescription which  accomplishes this   for 
the leading $1\ov \epsilon^2$ pole and, simultaneously,
 reproduces
 the finite part 
 of the logarithm of the exponentiated 4-gluon scattering amplitude \cite{bds}.


Let us consider
 applying  the  dimensional regularization  of \ci{am} 
 not at the level of the classical  superstring  action 
 but 
  only to the divergent quantities that arise during the calculation. 
  More precisely, using the $\epsilon=0$ solution for the minimal surface and
  taking into account 
   its homogeneity property, we  can first 
   factorize its area as was done in section  2.2 and 
   then  regularize the area  using the induced metric containing the 
   $1 \ov z^\eps$ factor.
   Demanding that 
   the coefficient of the second order pole in $\eps$ in 
    the expectation value of the
    single-cusp Wilson line matches that   in  the Sudakov form factor  will 
relate the dimensional continuation parameters on the gauge and the string sides. 
Applying   this relation to the 4-cusp Wilson loop 
we may  then reproduce the conjectured exponential 
structure of the  amplitude  \cite{ABDK,bds} to all orders in
string  perturbation theory.

As discussed in section 2, the homogeneity of the single-cusp
 solution and its relation to the spinning closed string solution imply that 
 the expectation value of the Wilson loop factor satisfies 
\be\la{wee}
\ln \langle W\rangle=\ln {\cal Z}=-f(\lambda)  \ A_2 
\ . 
\ee
In general,  $W$ may stand for all Wilson loops whose corresponding minimal
 surfaces are related by symmetries of \adss to the single-cusp Wilson loop
 surface.
  The difference between the minimal surfaces  will manifest itself 
  in the values of their  regularized areas  $A_2$.

An interesting observation is that while the 
expression above does not immediately reproduce the singular 
part of the 4-gluon scattering amplitude, 
it does,  however,  correctly  capture  its
 finite part. Indeed, the
  calculation  of the regularized area in \ci{am} yields
\bea
A_2^{\rm 4-cusp}
&=&2A_{{\rm div}, s}^{\rm string}+2A_{{\rm div}, t}^{\rm string}
-  \frac{1}{8}\left(\ln { s \ov t}\right)^2 + \const \ ,\\ 
A_{{\rm div}, s}^{\rm string}&=&\left[
\frac{1}{2 \epsilon_{\rm string}^2}
+\frac{1-\ln 2 }{4\epsilon_{\rm string}} \right]
\left(\frac{\mu^2}{|s|}\right)^{ \ha \epsilon_{\rm string}}
 \ . 
\la{py}\eea
Using it in equation \rf{wee} implies that the terms unrelated to the 
IR singularities are
\be
{\widetilde{\cal F}}_{1}=
\frac{f(\l)}{8} \ 
\left(\ln {s\ov t}\right)^2  +{\rm const}\ , 
\ee
which reproduces (up to a constant shift) the expected  structure 
of the finite part of the 4-gluon amplitude \ci{bds,am,dks}.
%

Assuming that there should indeed be a relation between the  null momentum 
Wilson loops and the gluon amplitudes, the observation above suggests that
we should aim at identifying $f(\l) \ A_{{\rm div}, s}^{\rm
string}$ with the logarithm of the Sudakov form factor, namely
\be
f(\lambda)\ A_{{\rm div}, s}( \epsilon_{\rm string},
\mu_{\rm string})=\frac{1}{8\epsilon_{\rm gauge}^2}f^{(-2)} (\td \l)
 +\frac{1}{4\epsilon_{\rm gauge}}g^{(-1)}(\td \l) \ , \ \ \ \ \ 
 \td \l \equiv 
\lambda \left( \frac{\mu_{\rm gauge}^{2}}{|s|}\right)^{\epsilon_{\rm gauge}}      
\label{ident}
\ee
This identification should hold up 
to   ${\cal O}(\epsilon_{\rm string})$     corrections. 
By comparing the
second 
order poles it is easy to see that they can be made to agree if we
choose\foot{This identification also leads to a successful comparison of
 the coefficient of the double-pole in the vev of the single-cusp 
 Wilson line and that of the Sudakov form factor.
  Indeed, the equation \rf{ea} implies that 
$$
-\ln \langle W_{\rm 1\,cusp}\rangle= 
\frac{
f(\lambda)}{\epsilon_{\rm string}^2}
+{\cal O}({1 \ov \epsilon_{\rm string}})=\frac{f^{(-2)}(\lambda)}{\epsilon_{\rm gauge}^2}
+{\cal O}({1\ov \epsilon_{\rm gauge}}) \ . 
$$}
\be
\epsilon_{\rm string}=\epsilon_{\rm
gauge}\sqrt{\frac{f(\lambda)}{f^{(-2)}(\lambda)}} \ .
\ee
It appears hard if not impossible, however,  to match the simple
 pole and the finite parts on the l.h.s. and r.h.s. of the equation \rf{ident}.
  To match the simple pole it seems necessary to demand that  
  $(\mu^2/s)_{\rm string}=(\mu^2/s)^{q(\lambda)}_{\rm gauge}$.
   Further 
ad hoc  adjustments appear to be needed in order to match the 
finite parts of  \rf{ident}.


\end{document}